\documentclass[12pt]{article}
\usepackage{epsfig}
\usepackage{mathtools}
\usepackage{tabulary}
\usepackage{array}
\usepackage[version=4]{mhchem}
\usepackage{float}
\usepackage[utf8x]{inputenc}
\usepackage{amsmath}
\usepackage{caption}
\usepackage{multirow}
\usepackage{geometry}
\usepackage{pdflscape}
\usepackage{booktabs}
\usepackage{amssymb}

\def\beq{\begin{equation}}
\def\eeq#1{\label{#1}\end{equation}}
\def\eeqn{\end{equation}}

\def\beqa{\begin{eqnarray}}
\def\eeqa#1{\label{#1}\end{eqnarray}}
\def\eeqan{\end{eqnarray}}

\let\bar=\overbar

\def\Dslash{\not{\hbox{\kern-4pt $D$}}}
\def\dslash{\not{\hbox{\kern-2pt $\del$}}}

\def\msb{{\bar{\ssstyle M \kern -1pt S}}}

\def\Title#1{\begin{center} {\Large {\bf #1} } \end{center}}

\begin{document}

\Title{Fractal Properties and Characterizations }

\bigskip\bigskip


\begin{raggedright}  

{\it John Hongguang Zhang\index{Zhang, J.}\\
Altamont Research\\
Altamont, NY12009}
\bigskip\bigskip
\end{raggedright}

{\bf Abstract:}  There are three important types of structural properties that remain unchanged under the structural transformation of condensed matter physics and chemistry. They are the properties that remain unchanged under the structural periodic transformation-periodic properties. The properties that remain unchanged under the structural multi scale transformation-fractal properties. The properties that remain unchanged under the structural continuous deformation transformation-topological properties. In this paper, we will describe some important methods used so far to characterize the fractal properties, including the theoretical method of calculating the fractal dimension, the renormalization group method, and the experimental method of measuring the fractal dimension. Multiscale fractal theory method, thermodynamic representation form and phase change of multiscale fractal, and wavelet transform of multiscale fractal. The  development of the fractal concept is briefly introduced: negative fractal dimension, complex fractal dimension and fractal space time. New concepts such as balanced and conserved universe,the wormholes connection to the whiteholes and blackholes for universes communication, quantum fractals,  platonic quantum fractals for a qubit, new manipulating fractal space time effects such as transformation function types, probabilities of measurement,manipulating codes,and hiding transformation functions are also discussed.  In addition, we will see the use of scale analysis theory to stimulate the elements on the fractal structure: the research on the dynamics of fractal structure and the corresponding computer simulation and experimental research. The novel applications of fractals in integrated circuits are also discussed in this paper.

\section{Introduction}

Many technological achievements in human history actually originate from imitation of nature. We often use "clever craftsmanship" to describe the ingenuity of a production. Therefore, studying the various growth patterns exhibited by the Great Nature is of great significance to our scientific and technological progress.We know that the traditional Euclidean geometry perfectly describes regular geometric shapes, while the classic Newtonian mechanics reveals the continuous change process of object dynamics. These brilliant achievements are the result of a high degree of generalization and abstraction of the real form, so they are often imperfect or even failed to accurately describe the real, discrete form. The exploration of mathematical structures other than Euclidean and Newtonian models began with modern Contor's set theory and Peano's "Space-filingcurves". Although these math masters at the time consider these problems and achievements belong to the categories of "abstract" and "beyond nature" such as cubist painting and atonal music, but in fact, human imagination always cannot escape the constraints of the natural environment on which they depend. People today find natural prototypes in what used to be called "morbid sets" or "morbid geometry". And further developed the ideas of Contor, Peano, Lebesgue, Hausdorff, Besscooitch Koch, and Sierpinski who have made outstanding contributions to the study of such structures, and established a new geometry to study natural forms.This new geometry is now called "Fractal geometry". It was developed by French American mathematician Benort B. Mandelbrot. In 1977, Mandelbrot published his first book: Fractals: Form, Chance and Dimension. Revised as The Fractal Geometry of Nature after five years~\cite{Mandelbrot}. 
In this book, Mandelbrot points out the scope of research on fractal geometry: Fractal geometry describes a class of irregular, fragmented patterns in nature ... The most useful fractals involve chance and both their regularities and their irregularities are statistical. Also, the shapes described here tend to be scaling, implying that the degree of their irregularity and/or fragmentation is identical at all scales.
Fractals with invariance of scale transformation as a characteristic of matter in nature have been recognized by science today. Therefore, the concept of fractal has attracted the attention of scientists working in different fields, and some progress has been made in introducing this concept into their respective fields. These fields currently include physics, chemistry, biomaterials, geology, seismic science, etc.~\cite{Pietronero1986,Avnir,Feder,Vicsek,Herrman}. In particular, the research on the dynamic processes on fractal bodies in recent years has greatly deepened people's understanding of the nature of fractal physics, showing its broad prospects for application in the field of physics.

\section{Physical Concept of Fractal }

\subsection{Fractal Concept}
The definition of fractals given by Mandelbrot in mathematics is related to the concepts of sets and topology: If a set is between the Hausdorff-Besicoritch dimension D and the topological dimension $D_T$ in the metric space and satisfies the relationship  $D>D_T$, then the set is called a fractal set.

In physics, the definition of self-similar fractal dimension is given by the following formula from a practical and clear perspective:

\begin{equation}
M(\lambda L)=\lambda^{D_f}M(L)
\label{equ:f1}
\end{equation}

Here $D_f$ is fractal dimension, $\lambda$ is scaling factor. Equation~\ref{equ:f1} \index{equ:f1} show the relationship between the lineality $L$ and the mass $M$ of an object. If the lineality of a fractal is transformed and its mass becomes the original $K$ times, its fractal dimension can be obtained by equation~\ref{equ:f1} \index{equ:f1} as following:

\begin{equation}
M(\lambda L)= KM(L)=\lambda^{D_f}M(L)
\label{equ:f2}
\end{equation}

\begin{equation}
D_f = \frac{ln K}{ln \lambda}
\label{equ:f3}
\end{equation}
equation~\ref{equ:f2} \index{equ:f2} and equation~\ref{equ:f3} \index{equ:f3}are the two basic formulas that are widely used to calculate the fractal dimension in applications. Many of the experimental methods used to calculate fractal dimension are developed based on it. Now we give two specific examples. In condensed matter physics, for a fractal growth group, if we replace the linear scale factor $\lambda$ in equation~\ref{equ:f2} \index{equ:f2} with the average radius of gyration $R$ of the group:
 
\begin{equation}
M \propto R^{D_f}
\label{equ:f4}
\end{equation}

If the group is embedded in d-dimensional Euclidean space, its "mass" density in Euclidean space is:

\begin{equation}
\rho \propto R^{D_f-d}
\label{equ:f5}
\end{equation}

equation~\ref{equ:f5} \index{equ:f5} is important to consider experimentally measuring the fractal dimension of fractal growth groups. 
Below we give another important equivalent expression of equation~\ref{equ:f3} \index{equ:f3} form from the perspective of measurement. From the perspective of measurement, to measure an area $S$, you can cover it with a small circle area with a radius $R$. The number of small circle area is,

\begin{equation}
N= \frac{S}{\pi R^2}\propto \frac{S}{R^2}
\label{equ:f6}
\end{equation}

Similarly, to determine a volume $V$, a small ball with a radius R can be used to fill it. The number of small balls required is:

\begin{equation}
N= \frac{V}{\frac{4}{3}\pi R^3}\propto \frac{V}{R^3}
\label{equ:f7}
\end{equation}

To generalize, for a fractal A with dimension $D_f$, the number of small balls needed

\begin{equation}
N \propto \frac{A}{R^{D_f}}
\label{equ:f8}
\end{equation}
equation~\ref{equ:f8} \index{equ:f8}is very important in practical applications. It is the basis of the fractal dimension determined by the box-counting method that is widely used at present~\cite{Russel,Frochling,Grrassberger,Greenside,Schaefer,Falconer1990}. The equivalence of equation~\ref{equ:f8} \index{equ:f8} and equation~\ref{equ:f3} \index{equ:f3} can be considered from two aspects. First, keep the radius $R$ of the small ball constant, and make the size of each dimension of the object enlarge $\lambda$. At this time, the object is enlarged $K$ times. Obviously the number of small balls becomes

\begin{equation}
N' \propto \frac{K A}{R^{D_f}}
\label{equ:f9}
\end{equation}

Secondly, we keep A unchanged, and reduce the radius of the ball by a factor of $\lambda$. At this time, the number of balls required must be 

\begin{equation}
N' \propto \frac{A}{( \,\frac{R}{\lambda}) \,^{D_f}}
\label{equ:f10}
\end{equation}

Comparing equation~\ref{equ:f9} \index{equ:f9} and equation~\ref{equ:f10} \index{equ:f10}, we get

\begin{equation}
K = \lambda^{D_f}
\label{equ:f11}
\end{equation}

\begin{equation}
D_f= \frac{lnK}{ln\lambda}
\label{equ:f12}
\end{equation}

From the above analysis, we can clearly see that the invariance of the scale transformation equation~\ref{equ:f1} \index{equ:f1} is the most fundamental characteristic of a fractal, and it is the fundamental starting point for handling all self-similar fractal problems.

\subsection{Definition of Various Fractal Dimensions}

Dimensions are important geometric parameters of space and objects. It is true that the mathematician Hausdorff has already defined the dimension of non-integer, but the concept of integer dimension has been inherited for a long time. It was only in the 1970s that many branches of the natural sciences successively introduced the concept of fractal dimensions in order to solve their own difficulties.

\subsubsection{Kolmogorov capacity dimension}

If $N(\varepsilon)$ is the minimum number of balls with diameter $\varepsilon$ (referred to as $\varepsilon$-ball) that can cover any point set (obviously, $N(\varepsilon)$ increases when $\varepsilon$ decreases). The capacity dimension of the point set is defined as~\cite{Besicvitch,Falconer1985,Grasbergerand1983A}:

\begin{equation}
D_0= \lim_{\varepsilon\to 0}{\frac{lnN(\varepsilon)}{ln(1/\varepsilon)}}
\label{equ:f13}
\end{equation}

\subsubsection{Information dimension}

It is not difficult to see that in the definition of the capacity dimension, there is a flaw: only the number of $\varepsilon$-spheres is considered in the definition,there is no difference in the number of points covered by each ball. A morbid case is that, according to this definition, the dimension of the set consisting of all rational numbers is 1, and the dimension of the set of real numbers is also 1. This is inconsistent with the spirit of real variable function theory. As an improvement, the definition of the information dimension is produced~\cite{Farmer1982B,Balatoni}:

\begin{equation}
D_1= \lim_{\varepsilon\to 0}{\frac{{\sum\limits_{i=1}^N }P_iln(1/P_i)}{ln(1/\varepsilon)}}
\label{equ:f14}
\end{equation}

Where $P_i$ is the probability of a point falling on the $i-th$ ball. When $P_i = 1 / N, D_1 = D_0$. It can be seen that the information dimension is a generalization of the capacity dimension. Because the information dimension involves the calculation of probability, it has a certain distance from the actual application, and the correlation dimension makes up for this deficiency.

\subsubsection{Correlation dimension}

In practical work, the correlation dimension is widely used. It is a dimension extracted from experimental data. Assume that the following data series were measured in the experiment:$(x_1,x_2,\cdots,x_i,\cdots, x_N)$ Here $N$ can be regarded as the total number of sample points, and $x$ is the value measured at the $i_{th}$ time. Since this is a set of time-related data, it is also called "time series". Take $(x_i,x_{i+\tau},\cdots, x_{i+(m-1)\tau})$ in this large series, so that you can get vectors in turn. Total $K=N-(m-1)\times\tau$ vectors. Use these vectors to support what is called an embedding space. The size of this space has a certain relationship with the dimension to be determined, that is, $m \geqslant 2D +1$ should be taken.The $\tau$ in the sequence $(x_i,x_{i+\tau},\cdots, x_{i+(m-1)\tau})$  is a time delay, which is also a parameter that is closely related to the physical content of the measurement signal.The correlation dimension is defined as~\cite{Grassberger1983B,Grassberger1986,Procaccia,Nicolis,Tsonis,Zhang1991}:

\begin{equation}
D_2 = \lim_{\varepsilon\to 0}{\frac{ln C(\varepsilon)}{ln(\varepsilon)}}    
\label{equ:f15}
\end{equation}

Here
$C(\varepsilon) =\frac{1}{K^2}\sum\limits_{i,j} \theta(\varepsilon-|x_i-x_j|) \\
\theta(\varepsilon-|x_i-x_j|)=\left.\{
  \begin{array}{lr}
    1, &   \varepsilon-|x_i-x_j| \geq 0\\
    0, &   \varepsilon-|x_i-x_j| < 0
  \end{array}
\right.$
Research shows that the size of the correlation dimension is limited by the total number of sample points~\cite{Eckmann}:

\begin{equation}
D_2 \leq 2logN/log1/\rho   
\label{equ:f16}
\end{equation}

\subsubsection{Generalized dimension}

Generalized dimension defined by Halsey can be writen as~\cite{Halsey}:

\begin{equation}
D_q = \lim_{\varepsilon\to 0}\lim_{q'\to q}{\frac{1}{q'-1}\frac{ln{\sum\limits_{i=1}^n }P_i^q}{ln\varepsilon}}    
\label{equ:f17}
\end{equation}

We will prove later: when $q=0,1,2$,  $D_q$ is the aforementioned Kolmogorov capacity dimension, information dimension and correlation dimension, respectively. Furthermore, the following properties can be proved: if $q>q'$, then $D_q<D_{q'}$.

\section{Experimental Method for Measurement of Fractal Dimensions}
In general, the experimental methods for determining the fractal dimension can be considered from the following five angles~\cite{Hentschel,Grassberger1984,Gao,Li}:The method of changing the degree of coarsening to find the dimension; the method of finding the dimension according to the measurement relationship; the method of finding the dimension according to the correlation function; the method of finding the dimension according to the distribution function. There are some areas that require common attention when using these methods. The first is the problem of the range and level of self-similarity, or the upper and lower limit of self-similarity. For example, although the shape of a cloud is a fractal with self-similar characteristics over a large scale, However, if the size of the earth is used as a benchmark, a cumulus cloud is nothing more than a point. If it is based on the size observed in a microscope, the cloud is just a collection of small water droplets, and it does not become self similar. Therefore, in the experiment to determine the fractal dimension,
There are certain limits on the self-similar range and hierarchy of objects. E. Hombogen pointed out in a review~\cite{Hombogen1989}: the range of self-similarity, the ratio of the maximum scale to the minimum scale is greater than one order of magnitude, and the self-similarity level $\geq$3

The second issue that needs attention is whether the same measurement value can be given when the same fractal object is measured using the different methods described above. There is no conclusive conclusion on this issue, and only specific cases can be analyzed.

Finally, there may be several fractal dimensions for complex fractals, that is, there are several linear scale regions on the double logarithmic curve~\cite{Hombogen1989,Long1991}.

\subsection{Method for determining dimensionality by changing degree of coarsening and method for determining dimensionality based on measurement relationship}

In previous Section, we gave the following formula from the perspective of measurement:

\begin{equation}
N(r) \propto r^{-D_f}
\label{equ:f18}
\end{equation}
Using formula ~\ref{equ:f18}, if we use a circle and a ball with radius $r$, or a line segment with side length $r$,  squares and cubes to have a basic figure with a characteristic length to approximate the fractal figure, the number of graphs $N(r)$ can determine the fractal dimension $D_f$. In this method, if we measure $N(r)$ to find $D_f$ by continuously changing the feature length $r$ of the basic figure and performing different levels of coarse viewing, it is called a method of changing the degree of coarse viewing to find the dimension. Typical box-counting methods measure fractal dimensions~\cite{Mandelbrot1984} are different from the above method, if we fix the size of $r$ unchanged, and "quantize" the fractal of the measured object, that is, decompose it into small islands, for each island, use the measurement relationship to determine the fractal dimension. it is called a method of finding the dimension based on the measurement relationship.This is typical of Mandelboot's "island" method for measuring fractal dimensions~\cite{Long1991, Mandelbrot1984}.

\subsection{Method of finding dimension based on correlation function}
The correlation function is one of the most basic statistics function. From this function type, the number of dimensions can also be obtained.
If the density at a certain coordinate $X$ of a random distribution in space is recorded as  $ \rho(X)$, then the correlation function $C(r)$ can be expressed as following~\cite{Avnir, Vicsek}:

\begin{equation}
C(r)\equiv< \rho(X)\rho(X+r)>
\label{equ:f19}
\end{equation}

Here $<...> $ means average. Depending on the situation, the average can be the overall average or the spatial average. If the distribution is equal in all directions, only the distance between two points $r=|\textbf{r}| $ function can be used to represent the correlation function.
As the function type of the correlation function $C(r)$, although the exponential type $e^{-r/r_0}$ and Gaussian type $e^{-r^2/2{r_0}^2}$ are usually compressed; considered as a model, they cannot be fractal. This is because they all have a characteristic distance $r_0$.  In $r_0 \ll r$ interval, related loss ratio is at $0<r<r_0$,  the decline in this interval is even more dramatic. In other words, when the distance between two points is less than $r_0$,  these two points strongly affect each other. But when the distance between the two points is greater than $r_0$, the two rooms hardly affect each other. Corresponding to this, when the distribution is fractal, the correlation function is power type. If it is a power type, there is no characteristic length, and the correlation is always attenuated by the same proportion. For example, suppose:

\begin{equation}
C(r)\propto r ^{-\alpha}
\label{equ:f20}
\end{equation}

can prove:

\begin{equation}
\alpha=d-D_f
\label{equ:f21}
\end{equation}

\subsection{The method of finding the dimension according to the distribution function}

If you only look at the photos of moon pits of various sizes on the lunar photos, the scale bar is completely invisible. If the diameter of the moon pit on the photo is 1000 $km$, it will feel quite large. If it is 50 $cm$, it will only make people feel that they are so small, which will not particularly make people feel unnatural. The size distribution of the moon pit has no characteristic length.When considering this size distribution, the scoring dimension can be obtained from the type of its distribution function. Let the diameter of the lunar crater be $r$, and the probability of the existence of a lunar crater with a diameter greater than  $r$ as $P(r)$. If the probability density of the diameter is $p(r)$, then~\cite{Gao} :

\begin{equation}
P(r)=\int_r^\infty p(s)\mathrm{d}s
\label{equ:f22}
\end{equation}

The ability to change the scale of photos and pictures means that it can correspond to the transformation  into $r \rightarrow \lambda r $. Therefore, if you want to change the scale without changing the distribution type, for any input $\lambda > 0$

\begin{equation}
P(r)\propto P(\lambda r)
\label{equ:f23}
\end{equation}

This relationship must be established. The $r$ function type that can always satisfy the above formula is limited to the following power types:

\begin{equation}
P(r)\propto r^{-D}
\label{equ:f24}
\end{equation}

It is not difficult to understand that the power index $D$ can give the fractal dimension of the distribution if it can be considered as follows. Consider using a coarse visualization of the moon pits that cannot see when less than $r$, then the number of visible moon pits is proportional to $P(r)$, because $N(r)$ is proportional to $P(r)$, the $D$ that appears here is consistent with the definition of $D$ given by  ~\ref{equ:f18}

\subsection{Method of finding dimension according to spectrum}
When investigating the statistical properties of random variables in space or time based on observations, it is often easier to derive a spectrum that uses wavenumber decomposition. For example, when a random time variation $ X(t)$ is given, the spectrum $S(f)$ with the vibration number $f $, based on Fourier transform  $x(f)\equiv \int e^{ift}x(t)\mathrm{d}t $, can be defined as:

\begin{equation}
S(f)\equiv|x(f)|^2
\label{equ:f25}
\end{equation}

If the change changes by a certain constant, $S(f)$ is consistent with the Fourier transform of the correlation function. If the variation is converted into an electrical signal, then through a filter, the amount proportional to the power spectrum $S(f)$ can be obtained. Whether a change has fractal characteristics, also available from judging from the "according to the degree of coarsening" introduced earlier. From the point of view of the spectrum, the so-called change of the coarseness of observation is to improve the cut-off frequency. The cut-off frequency mentioned here refers to the cut-off frequency of the finer vibration components. Therefore, if a change has a fractal characteristic, it is equivalent to saying that even if the cut-off frequency $f_c$ is changed, the shape of the spectrum is not changed. This is also equivalent to that the shape of the transformed spectrum does not change even if the observation scale is transformed. The spectrum $S(f)$ with this property is limited to the following:

\begin{equation}
S(f)\propto f^{-\beta}
\label{equ:f26}
\end{equation}

If the spectrum is of this power type, it can be proved that the $\beta$ and fractal dimension D of the change curve graph satisfies:

\begin{equation}
\beta=5-2D
\label{equ:f27}
\end{equation}

When considering the curves of terrain and solid surface, the following expansion can be made. Assuming that the spectrum of the cross section obtained by cutting a curved surface with a plane is $S(f)$, the fractal dimension of the ground surface $D (2 <D <3)$ can satisfy the following formula:
\begin{equation}
\beta=7-2D
\label{equ:f28}
\end{equation}
It has been assumed here that the variation of the curved surface is equal.

\section{Renormalization Group Method for Calculation of Fractal Dimensions  }
The purpose of the renormalization group is to change the degree of coarseness in the observation to quantitatively obtain the change in physical quantity, so it is a powerful mathematical tool to describe the invariance of the scale. As mentioned earlier, the fundamental feature of fractal is that it has self-similarity under scale transformation or scale transformation invariance, so it is more appropriate to use the renormalization group to solve the fractal problem.
If the physical quantity measured on the basis of coarsening at a certain scale is recorded as $p$, when using a large-scale coarsening that is twice larger than this scale, this physical quantity is recorded as $p'$. Using the appropriate coarse-view transformation $f_2$, we can express $p'$ and the original $p$ as follows:

\begin{equation}
p'=f_2(p)
\label{equ:f29}
\end{equation}

Among them, the lower angle $2$ of $f$ represents twice the coarsening. If the scale of the coarsening is enlarged twice, then the following relationship holds:

\begin{equation}
P''=f_2(p')=f_2 \bullet f_2(p)=f_4(p)
\label{equ:f30}
\end{equation}

If we turn this formula into a generalized relationship, then we know that the transformation $f$ has the following properties:

\begin{equation}
f_a \bullet f_b=f_{ab}
\label{equ:f31}
\end{equation}

\begin{equation}
f_1=1
\label{equ:f32}
\end{equation}

Where $1$ represents the identity transformation. The transformation $f$  generally does not have the inverse transformation $f^{-1}$ . This is because in a given state, you can always look at it roughly. But on the contrary, even if the coarse-viewed state is predetermined, its original state cannot be restored. Mathematically, the transformation with this property is called semigroup. This is because the transformation that uses coarse-viewing is physically called renormalization. It stands to reason that it is correct to call this $f$-transformation a renormalized semigroup. But now it is used to the name of renormalization group.

Roughly speaking, there are two types of random fractals: "medium" has a random fractal structure (such as percolation group) and "fluid" has a random fractal structure (such as fractal growth group). The renormalization group method can be used to get the fractal dimension of these two random fractals.

\subsection{Using Renormalization Group to Find Fractal Dimension with Random Fractal Structure}
Taking the percolation problem as an example, consider the state where the metal is distributed in a random state on the lattice points of the two-dimensional square grid, and take the probability of metal occupation as the physical quantity $p'$.

\begin{figure}[H]
\begin{center}
\epsfig{file=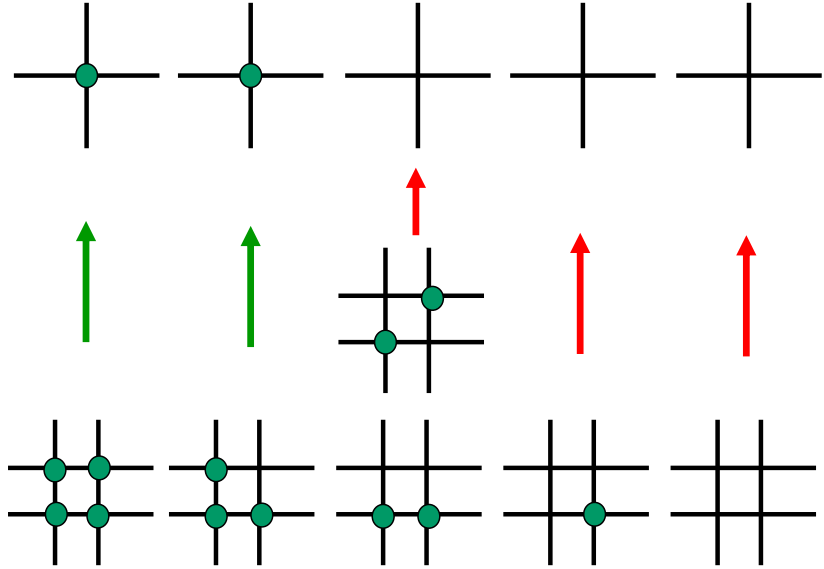,height=4.8in}
\caption{Coarsening $2 \times 2$ grid points on a hypothetical grid point}
\label{fig:figure1}
\end{center}
\end{figure}

Now let's explain the problem of coarsening $2 \times 2$ grid points on a hypothetical grid point. Call this new grid super grid. To make a super grid, the $2 \times 2$ grid points after roughening are called blocks. When there are metals in the four points in the block, it can be considered that the super-lattice point after roughening this block also has metal. When there is metal on the three lattice points in the block, because the block can pass electricity in both the vertical and horizontal directions, it can also be considered that there is metal on the super lattice point. However, if the number of points in the block is less than two, the block cannot pass electricity at least in the vertical or horizontal direction. Therefore, this time should correspond to the state when there is no metal at the super lattice point (Figure~\ref{fig:figure1}). If the metal occupancy rate of the super lattice point metal is assumed to be $p'$, then the following equation hold:

\begin{equation}
p'=f_2(p)=p^4+4p^3(1-p)
\label{equ:f33}
\end{equation}

The first item indicates the state when metal is present at all four points within the block, and the second item indicates the state when metal is present at all three points. Because the transformation $f$ is determined by using this formula, as long as the nature of  $f $ is studied later, the phase change should be parsed.
As discussed earlier, the critical point $p_c$ is the invariant point after transformation $f$, that is, the fixed point. If the fixed point is assumed to be $p_n$, Then according to equation ~\ref{equ:f1} \index{equ:f1} , the following equation holds, namely:

\begin{equation}
p_n=p_n^4+4p_n^3(1-p_n)
\label{equ:f34}
\end{equation}

And $p_n$ can be obtained:

\begin{equation}
p_n=0,1, \frac{1\pm\sqrt{13}}{6}\approx-0.434, 0.768
\label{equ:f35}
\end{equation}

Among them, because $p$ represents probability, so $p_c = -0. 434$ should be excluded. If $p_n = 0$ means no no metal at all, $p_n = 1$ means all metal,  then $p_c = 0.768$ is the critical state with metal. The fractal dimension of the percolation group at the critical point will be required below. If the super-lattice point is metal, there are three or four points within the block. The expected value of the number of metal grid points in this block is:

\begin{equation}
N_c=\{4\bullet p_c^4 +3\bullet4 \bullet p_c^3(1-p_c)\}/p_c\approx 3.45
\label{equ:f36}
\end{equation}

It is removed by $p_c$ because this expected value refers to the expected value under the condition that the superlattice is metal. If with super lattice, In contrast, the grid interval is $1/2$, that is to say, in the super grid, since the observation unit length is set to $1/2$, therefore, of the average $N_c$ metal points, only one metal point should be visible. If the above relationship is changed to a general relationship,then when the observed unit length is $1/b$ times, the number of metal points $N_c(b)$ that can be seen is:

\begin{equation}
N_c(b)=b^{-D}
\label{equ:f37}
\end{equation}

The above formula is essentially the measurement relationship of ~\ref{equ:f18}. So for the fractal group of percolation:

\begin{equation}
D=\frac{logN_c}{log b}=\frac{log3.45}{log 2}\approx 1.79
\label{equ:f38}
\end{equation}

$D$ is the fractal dimension of the percolation fractal group, which is completely consistent with the value obtained by simulation. It is also approximate to the experimental value of $1.9$. This confirms the superiority of the renormalization group.

\subsection{Finding the fractal dimension of fractal growth using renormalization group: dynamic renormalization group method for fractal growth}

\begin{figure}[H]
\begin{center}
\epsfig{file=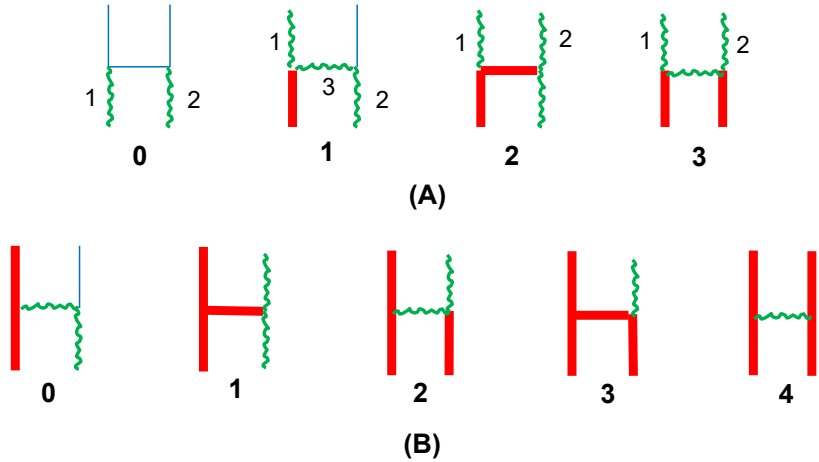,height=4.8in}
\caption{Three categories of keys on the two-dimensional square lattice}
\label{fig:figure2}
\end{center}
\end{figure}

Taking DLA fractal growth as an example~\cite{Wang1989,Huang,Nagatani1991}: consider the dynamic renormalization group of a two-dimensional square lattice. First, divide the keys on the lattice into three categories: (i) group keys: they form a fractal group (represented by a thick solid red line in Figure ~\ref{fig:figure2}) (ii) growth keys: they are adjacent to the surface of the fractal group. The key is that can continue the growth of the group (represented by the wavy green line) (iii) The empty key: the key that is not occupied by the fractal group or adjacent to the group (represented by the thin blue line) The calculation assumes that the horizontal and vertical directions are decoupled.

The theory proves that this assumption is applicable to isotropic growth groups and does not affect its scaling behavior. Then do kadanoff transformation and stipulate that the vertical (horizontal) direction from bottom to top (from left to right) in the original lattice is occupied by group keys. 
The breakpoint matrix is ​​renormalized to a group bond;  when it cannot be completely occupied by the group bond, but it contains a growth bond, it is renormalized to a growth bond; It is stipulated that when there is only an empty key, it is renormalized into an empty key. According to these rules, for a transformation $k = 2$, the four configurations in Figure~\ref{fig:figure2}A can be renormalized into a vertical growth key. From the electrostatic analogy method, it can be assumed that the growth probability $p$ of the group bond at each point is proportional to the local electric field $E$ at that point, ie $p \sim  E$~\cite{Nagatani1987A,Nagatani1987B}. The keys have the same conductivity $\sigma $. Taking configuration 1 in Figure ~\ref{fig:figure2} as an example, the voltage on growth key 1 is $ E_{11} = l$, and the voltages on growth keys 2 and 3 should be equal. $ E_{12} = E_{13}$, using Kirchhoff's law:

\begin{equation}
\sigma (1-E_{12})=2 \sigma E_{12}
\label{equ:f39}
\end{equation}

$ E_{12}=E_{13}=\frac{1}{3}$

From this, the growth probability of growth keys 1,2,3 in configuration l can be obtained as:

\begin{equation}
p_{11}=\frac{E_{11}}{E_{11}+E_{12}+E_{13}}=0.6
\label{equ:f40}
\end{equation}

\begin{equation}
p_{12}=p_{13}= \frac{E_{12}}{E_{11}+E_{12}+E_{13}}=0.2
\label{equ:f41}
\end{equation}

Similarly, the growth probability of other configurations can be obtained as

\begin{equation}
p_{01}=p_{02}= p_{21}=p_{22}=p_{31}=p_{32}=0.5
\label{equ:f42}
\end{equation}

Next is to ask for the statistical weight of the appearance of various configurations. It can be seen from the Figure~\ref{fig:figure2} that configuration (1) is grown by the growth key 1 of configuration (0), which is symmetrical to that of configuration (0) Key 2 can also generate configuration (2). So the statistical weight of configuration (1) can be written as:

$c_1=c_0p_{01}+c_0p_{02}$

Using the above rules and normalization conditions, the statistical weights of the four configurations are

\begin{equation}
c_0=c_1=\frac{5}{12}=0.4167; c_2=c_3=\frac{1}{12}=0.0833
\label{equ:f43}
\end{equation}

Figure (~\ref{fig:figure2}B) shows the various configurations renormalized into a vertical group key. Similarly, their statistical weights $W_{\alpha}$ are expressed as

\begin{equation}
W_0=0.6c, W_1=W_2=W_3=W_4=0.5c
\label{equ:f44}
\end{equation}

According to the scale invariance, we expect that both $c_{\alpha}$ and $W_{\alpha}$ will remain unchanged under the renormalization transformation. It is easy to write the recursive relationship between the group mass $M_n$ and the surface mass $m_n$.

\begin{equation}
\begin{aligned}
m_n &=(2c_0+ 2c_1 +3c_2+ 2c_3)m_{n-1}+(c_1 +c_2+ 2c_3)M_{n-1} \\
&=\alpha_1m_{n-1}+ \beta_1 M_{n-1}
\end{aligned}
\label{equ:f45}
\end{equation}

\begin{equation}
\begin{aligned}
M_n &=(2W_0+ 2W_1 +3W_2+ 3W_3+4W_4)M_{n-1}+(W_0 +2W_1+ W_2+W_3)M_{n-1}\\
&=\alpha_2M_{n-1}+ \beta_2 m_{n-1}
\end{aligned}
\label{equ:f46}
\end{equation}

Substituting the previous values ​​of $c_{\alpha}$ and $W_{\alpha}$, we get

$ m_n=2.0833m_{n-1}+0.667M_{n-1}$ and  $M_n=2.769M_{n-1} + m_{n-1}$

The surface quality and the recursive form of fractal groups have the following general form (also known as the transformation matrix of the renormalization group)

\begin{equation}
 \begin{bmatrix}
       m \\[0.3em]
       M
     \end{bmatrix}_n=\begin{bmatrix}
       \alpha_1 & \beta_1   \\[0.3em]
       \beta_2 & \alpha_2 
     \end{bmatrix}\begin{bmatrix}
       m \\[0.3em]
       M
     \end{bmatrix}_{n-1}
\label{equ:f47}
\end{equation}

For the DLA growth group, as the computer simulation results, most of the group bonds are surface growth bonds, so the fractal dimension of the group is the same as the surface fractal dimension. This means that there is the following asymptotic scaling behavior:

\begin{equation}
\lim_{n\to \infty}{M_n/M_{n-1}}= \lim_{n\to \infty}{m_n/m_{n-1}}    
\label{equ:f48}
\end{equation}

Near the stable fixed point, the RG transformation matrix ~\ref{equ:f47} has the following two eigenvalues

\begin{equation}
 \lambda_\pm=\{(\alpha_1+\alpha_2)\pm [ \,(\alpha_2-\alpha_1)^2+4\beta_1\beta_2] \,^{1/2} \}/2  
\label{equ:f49}
\end{equation}

Put the $\alpha_i$ and $\beta_i$value in, we get $\lambda_+ = 3.311$ and $\lambda_- = 1.540$.
 From equations ~\ref{equ:f45} and ~\ref{equ:f46}, we obtain:

\begin{equation}
\frac{m_n}{m_{n-1}} = \alpha_1+ \beta_1\frac{M_{n-1}}{m_{n-1}} 
\label{equ:f50}
\end{equation}

\begin{equation}
\frac{M_n}{M_{n-1}} = \alpha_2+ \beta_2\frac{m_{n-1}}{M_{n-1}} 
\label{equ:f51}
\end{equation}

If we let

\begin{equation}
\lim_{n\to \infty}{\frac{m_{n-1}}{M_{n-1}}}= R 
\label{equ:f52}
\end{equation}

\begin{equation}
\alpha_2+\beta_1\frac{1}{R} = \alpha_2+ \beta_2R 
\label{equ:f53}
\end{equation}

The solution of equation ~\ref{equ:f53} is:

\begin{equation}
R= \frac{(\alpha_1-\alpha_2)+\sqrt{(\alpha_2-\alpha_1)^2+4\beta_1\beta_2}}{2\beta_2}
\label{equ:f54}
\end{equation}

Take $R$ into equation  ~\ref{equ:f51}, we obtain:

\begin{equation}
\frac{M_n}{M_{n-1}} = \alpha_2+ \beta_2R= \lambda_+
\label{equ:f55}
\end{equation}

From the basic characteristics of fractal equation ~\ref{equ:f1}

\begin{equation}
M(b L)=b^{D_f}M_{n-1}(L)
\label{equ:f56}
\end{equation}

\begin{equation}
b^{D_f}=\frac{M_n}{M_{n-1}}=\alpha_2+ \beta_2R= \lambda_+
\label{equ:f57}
\end{equation}

\begin{equation}
D_f=\frac{ln\lambda_+}{lnb}=1.727
\label{equ:f58}
\end{equation}

The above calculated values ​​are in good agreement with experimental values ​​and computer simulation values. 
Huang and Nagatani~\cite{Huang,Nagatani1991,Nagatani1987A,Nagatani1987B}  promoted the above method from two aspects. On the one hand, it is applied to the $\eta $ model, that is, assuming the relationship between the growth probability $p$ of the group surface and the local electric field $E$ is $p \sim E^{\eta}$. Here $\eta$ is a parameter, when $\eta=1$, it is DLA model.  Similarly, for a two-dimensional square lattice, find:

\begin{equation}
\begin{aligned}
c_0 =c_1= \frac{2+3^\eta}{2(3+3^\eta)}, c_2 =c_3= \frac{1}{2(3+3^\eta)} \\
W_0 = \frac{3^\eta}{4+3^{\eta+1}}, W_1 =W_2=W_3=W_4= \frac{2+3^\eta}{2(4+3^{\eta+1})}
 \end{aligned}
\label{equ:f59}
\end{equation}

When $\eta=2$, $\lambda_+ = 3.2093$ , $D=1.68$, when $\eta \to \infty $, $\lambda_+ = 3.115$ , $D=1.64$, from this results, we can see that parameter $\eta$ has less effect on the dimension calculation.

On the other hand, it is extended to any dimensional space. From the previous description, it can be seen that from 2D to 3D, the calculation of the configuration after the cell transformation will be very complicated. To this end, a theoretical result $D(q=\infty ) = D-1$ is used, from the definition of $D_q$, we have $D(q=\infty ) = -\frac{ln p_{max}}{ln2}$,  so that for a configuration, only the maximum growth probability of all group keys is required, and it is not necessary to find the growth probability of all group keys. This simplifies the calculation. Using this idea, the general formula for any dimension is :

\begin{equation}
p_{max}=\frac{1}{1+2c_{d-1}^1 \varphi_1+\sum\limits_{k=2}^{d-1}c_{d-1}^k\varphi_k}
\label{equ:f60}
\end{equation}

\begin{equation}
D=1+\frac{ln(1+2c_{d-1}^1 \varphi_1+\sum\limits_{k=2}^{d-1}c_{d-1}^k\varphi_k)}{ln2}
\label{equ:f61}
\end{equation}

Where $\varphi_k$ represents the potential of the $i-th$ bond. The calculation result is listed in Table~\ref{tab:T1}:

\begin{table}[htb]
\begin{center}

\begin{tabular}{p{3cm}p{3cm}p{5cm}} \hline 
d &  $D=1+D_{\infty} $  & D (simulation value) \\ \hline 
2 &  1.737   & $1.75 \pm 0.004$ \\
3 &  2.515   & $2.495 \pm 0.005$ \\
4 &  3.341   & 3.40 \\

 \hline
\end{tabular}
\caption{Comparison of the theoretical results and the computer simulation results}
\label{tab:T1}
\end{center}
\end{table}

The agreement between the theoretical results and the computer simulation results ~\cite{Tolman} in  Table~\ref{tab:T1} shows that this simplified calculation method can basically reflect the structural properties of fractals. Above we introduced the calculation method of the renormalization group theory of the fractal dimension of two large systems with random fractal structure in condensed matter physics. However, the real power of the renormalization group is that it can analytically find the criticality of phase transition and critical phenomena index, through the critical index we can also find the fractal dimension of the fractal structure in the phase transition and critical phenomenon.

\section{Multi Scale Fractal Theoretical Methods }
\subsection{Basic formula of multi-scale fractal theory}
In-depth studies of the fractal structure in recent years have shown that although the scale relationship in the sense of fractal gives a certain value-a fractal dimension, such a fractal dimension, in addition to marking the self-similar structural law of the structure, it does not fully reveal the dynamic characteristics of the corresponding structure.

The disadvantage of using a single fractal dimension to describe the structure formed after a more complicated nonlinear dynamic evolution process is that it is too general. We know that in most physical phenomena, the behavior of the system mainly depends on the spatial distribution of a certain physical quantity (usually a scalar quantity, such as concentration, electric potential, probability, etc.). Therefore, if one considers the physical nature of the fractal body involved in the fractal concept
There should be a similar spatial distribution of a certain amount in the phenomenon. The theory of multi-scale fractal is to study this behavior of fractal. What it discusses is the probability distribution of a certain parameter~\cite{Amitrano,Halsey1986,Turkevich,Feigenbaum1986A}.

Based on the well-known Euclidean geometry, the distribution functions and derivatives of all physical quantities on it are relatively smooth. In other words, the local exponential behavior of the distribution function (the corresponding scale exponent is called singularity) is very few. It doesn't even exist at all. However, for a fractal, such singularity ​​can be dispersed within a range, which can constitute a spectrum.

Generally, when we discuss the distribution of a physical quantity on a geometric structure, we divide the geometric structure into several small parts, such as $N$ parts ($N$ can be very large) and the $i-th$ part $(i = l, 2, , N)$ is represented by its linearity $l_i$. If these areas are small enough, the distribution of the physical quantity in this area can be regarded as uniform. Therefore, in any small area, the relationship between the physical quantity $P_i$ and its linearity is:

\begin{equation}
P_i \sim l_i^{\alpha_i}
\label{equ:f62}
\end{equation}

Where $\alpha_i = d$ is the Euclidean dimension. This exponential relationship is very common in physical phenomena, especially in critical phenomena, the exponential scaling formula is the most important physical law characterizing critical behavior. Taking the droplet theory of the Ising model as an example, in the region of size $l_i$, the magnetization intensity $M_i$ can be expressed as:

\begin{equation}
M_i \sim l_i^y
\label{equ:f63}
\end{equation}

Here $y$ is the standard critical scale index. In the critical phenomenon, such critical exponent is usually very small and small enough to establish a simple scaling relationship~\cite{Feigenbaum1986A,Ma,Widom,Kadanoff1967,Frieda}. But the nonlinear dynamic problem is similar to equation ~\ref{equ:f63}.  The scaling formulas are often in a series. Therefore, we can arrange these scaling formulas in a sequence~\cite{Kadanoff1966}:

\begin{equation}
(M_i)^q \sim l_q^y,       q=1,2,3\dots
\label{equ:f64}
\end{equation}

For the evolution process of a fractal, such as the growth of DLA group, it mainly depends on its “active area”, that is, its growth front part~\cite{Stanley}. In a real growth process, the local conditions of these active areas are not the same. Therefore, we can imagine that if the fractal geometry such as DLA group is divided into several small areas $l_i$, and the growth probability $P_i$ in each small area also has a scale relationship, then the scale index in these small areas can take a series of values:

\begin{equation}
P_i^q \sim l_i^{\alpha_q}
\label{equ:f65}
\end{equation}

If the sizes of the areas we divide are all the same $l_i$, and the density of the singularity $\alpha $ is $ \rho (\alpha)$, we further assume that $\alpha $ is in the area $\alpha'$ to $\alpha'+d\alpha'$, the number of times a certain value is taken up is:

\begin{equation}
d\alpha'\rho (\alpha')l^{-f(\alpha')}
\label{equ:f66}
\end{equation}

Here $f(\alpha) $ is an introduced continuous function, which reflects the number of times $\alpha $ is taken on a certain subset. Such an assumption is equivalent to a density. In the above two assumptions, equation ~\ref{equ:f65} is more universal, at least analytically, it can be proved that this scale expression exists. The main significance of the hypothesis expressed in formula ~\ref{equ:f66} is that it expresses the density of the singular value a as a continuous function $f(\alpha)$ .

Before linking $f(\alpha) $  to observable quantities, we first introduce the generalized dimension $D_q$. The dimension was introduced  with reference to the definition of Renyi information dimension when studying strange attractors, and its definition formula is:

\begin{equation}
D_q=\frac{1}{q-1} \lim_{l\to 0}\frac{logX(q)}{log l}
\label{equ:f67}
\end{equation}

where 

\begin{equation}
X(q)=\sum\limits_{i}(P_i)^q
\label{equ:f68}
\end{equation}

From equation ~\ref{equ:f67}, we can see that the subsets with different scale indices can be distinguished by the change of $q$ value. When $q = 0$

\begin{equation}
D_0=\lim_{l\to 0}\frac{lnN(l)}{ln1/l}
\label{equ:f69}
\end{equation}

This is the aforementioned Kormogorov capacity dimension

When $q=1$, let $P_i^q = P_iP_i^{q-1}= P_iexp [(q-1)lnP_i]$, when $q \to 0 $, from L.Hosptals rule, $exp [(q-1)lnP_i]= 1+(q-1)lnP_i$, and thus $ln\sum\limits_{i}(P_i)^q \to ln[1+(q-1)\sum\limits_{i}(P_i)\approx (q-1)\sum\limits_{i}(P_i)lnP_i]$, since $\sum\limits_{i}(P_i)=1 $, put into equation ~\ref{equ:f67}, we obtain:

\begin{equation}
D_1=\lim_{l\to 0}\frac{\sum\limits_{i}P_ilnP_i}{lnl}
\label{equ:f70}
\end{equation}

This is information fractal dimension
  
When q=2, equation ~\ref{equ:f67} becomes the relevant dimension:

\begin{equation}
D_2=\lim_{l\to 0}\frac{ln\sum\limits_{i}P_i^2}{logl}
\label{equ:f71}
\end{equation}

Therefore, The generalized dimension $D_q$ actually includes all dimensions involved in the self-similar fractal theory, and expands the connotation of the fractal theory.

In order to connect $D_q$ to the local scale characteristics, substituting equations ~\ref{equ:f65} and ~\ref{equ:f66} into equation ~\ref{equ:f68}, the moment $X(q)$ can be written as~\cite{Halsey}:

\begin{equation}
X(q)=\int \mathrm{d}\alpha'\rho (\alpha')l^{-f(\alpha')}l^{\alpha'_q}
\label{equ:f72}
\end{equation}

Let $\rho (\alpha'\neq 0)$, since $l$ is very small, the moment $X(q)$ only contributes the most when $q\alpha'-f(\alpha') $ takes a minimum value. Since $\alpha$ is arranged according to $q$, the condition for $q\alpha'-f(\alpha') $  to take the minimum value is

\begin{equation}
\left.\frac{\mathrm d}{\mathrm d \alpha'} [q\alpha'-f(\alpha')]|_{\alpha'=\alpha(q)}\right.=0
\label{equ:f73}
\end{equation}

\begin{equation}
\left.\frac{\mathrm d^2}{\mathrm d (\alpha')^2} [q\alpha'-f(\alpha')]|_{\alpha'=\alpha(q)}\right.>0
\label{equ:f74}
\end{equation}

And from equations ~\ref{equ:f67}, ~\ref{equ:f72},~\ref{equ:f73} and ~\ref{equ:f74}, we have:

\begin{equation}
D_q=\frac{1}{q-1} [q\alpha(q)-f(\alpha(q))]
\label{equ:f75}
\end{equation}

\begin{equation}
\alpha(q)=\left.\frac{\mathrm d}{\mathrm d q} [(q-1)D_q]\right.
\label{equ:f76}
\end{equation}

Therefore, if we measure the value of $D_q$, we can get $\alpha(q) $ from equation ~\ref{equ:f76}  and then find $f(\alpha(q)) $ from equation ~\ref{equ:f75}, otherwise, if we know $f(\alpha(q)) $ , then the $D_q$ value can be calculated by the above formula. In the experiment, $D_q$ is usually measured before calculating $f(\alpha(q))$ , because $D_q$ is easy to measure.

From equations~\ref{equ:f73} and ~\ref{equ:f74}, we can see the geometric characteristics of the $f(\alpha) $ curve ~\cite{Grassberger,Cates,Ball,Cawley,Huang1991,Platt,Hakasson}. Since $\frac{\partial f(\alpha)}{\partial \alpha} = -q, \frac{\partial^2 f(\alpha)}{\partial \alpha^2 < 0}  $, the $f(a)$ curve is convex to the $\alpha $ axis. The maximum value is obtained at $q =0$, It has an infinite slope at $q = \pm \infty$. When $q = l$, it is determined by equations ~\ref{equ:f75} and ~\ref{equ:f70}: $f(\alpha(1)) = \alpha(1) = D(1) \leqslant l$. These results above do not involve specific problems, so these geometric features of the $f(\alpha)$ spectrum are universal.

The above formulas are widely adopted and used in multi-scale fractal research. They constitute the main content of multi-scale fractal theory. The physical ideas contained in it reflect the evolution characteristics of complex structures produced by a large class of nonlinear dynamic systems. Therefore, multi-scale fractal theory is of great significance for understanding the formation of these structures and their dynamic roots.

\subsection{Thermodynamics and phase transition of multi-scale fractal}

The method of multi-scale fractal theory dividing large scale into many small scales is similar to the method of dividing the system into many macroscopically small and microscopically large subsystems in thermodynamics. We know that all the physical quantities in thermodynamics can be derived from the partition function. This makes us think that for multi-scale fractals, we can also start with a properly defined partition function and introduce a set of thermodynamic forms such as entropy and free energy. It will provide a new way to comprehensively describe and analyze the characteristics of multi-scale fractal. But it should be emphasized that this thermodynamic description is mainly formal, and what it reveals is still mainly the geometric characteristics of the body.

\subsubsection{The fractal entropy and free energy function}

We only study the topological aspects of fractal here, consider a systematically divided set, assuming that at step $n$, the set is divided into $N (n)$ balls, and at step $n+1$, each of these balls is divided a certain number of balls, there are $N (n + 1)$ balls. Suppose that in the $n-th$ division, the diameter $ l_i$ of the $i-th$ ball has a scale index $ \varepsilon_i $, which can be written as:

\begin{equation}
l_i=exp(-n\varepsilon_i)
\label{equ:f77}
\end{equation}

When $n$ becomes larger, $ l_i$ tends to zero, but $ \varepsilon_i $ takes a non-zero finite value. Here $ \varepsilon_i $ has the meaning of single particle "energy" in thermodynamic statistical physics. Let $\Omega(\varepsilon)d\varepsilon  $ be the number of balls whose scale index is between $\varepsilon_i$ and  $\varepsilon_i + d\varepsilon_i$ ,corresponding to the Boltzmann relationship in thermodynamics, we can
Definition $\Omega(\varepsilon)$ has the following scale form:

\begin{equation}
\Omega(\varepsilon)=exp[nS(E)]
\label{equ:f78}
\end{equation}

Here we call $S(E)$ as the entropy function of each step. Equation~\ref{equ:f78} can be regarded as the basic properties of fractal sets.

Introduce the partition function~\cite{Kohmoto}:

\begin{equation}
Z(\beta)=\sum\limits_{i=1}^{N(n)}l_i^\beta=\sum\limits_{i=1}^{N(n)}exp(-\beta n\varepsilon_i)
\label{equ:f79}
\end{equation}

Free energy defined as:

\begin{equation}
F(\beta)=\frac{1}{n}lnZ(\beta)=\frac{1}{n}ln\sum\limits_{i=1}^{N(n)}exp(-\beta n\varepsilon_i)
\label{equ:f80}
\end{equation}

Equation~\ref{equ:f79} integral form of $\varepsilon$:

\begin{equation}
Z(\beta)=\int \mathrm{d}\varepsilon\Omega exp(-\beta n\varepsilon)=\int \mathrm{d}\varepsilon   exp \{n[S(E)-\beta \varepsilon ]\}
\label{equ:f81}
\end{equation}

When $n$ is large, the above integral is dominated by the maximum value of $\varepsilon =\overline{\varepsilon}, S(E)-\beta \varepsilon $, thus

\begin{equation}
\left.\frac{\mathrm d S}{\mathrm d \varepsilon  } |_{\varepsilon=\overline{\varepsilon}}\right.=\beta
\label{equ:f82}
\end{equation}

From equations~\ref{equ:f80} and ~\ref{equ:f81},  free energy:

\begin{equation}
F(\beta)=S(\overline{\varepsilon})-\beta \overline{\varepsilon}
\label{equ:f83}
\end{equation}

It is Legendre transformation. From this we can see that $\beta$ and $F(\beta)$can be obtained from the entropy function. In addition, from equations~\ref{equ:f80} and ~\ref{equ:f83} we can get:

\begin{equation}
\overline{\varepsilon}=\left.\frac{\mathrm d F(\beta)}{\mathrm d \varepsilon }\right.=\sum\limits_{i=1}^{N(n)}exp(-\beta n\varepsilon_i)/Z(\beta)
\label{equ:f84}
\end{equation}

The above formula shows: $\varepsilon$ is the average value of the probability distribution of $\varepsilon_i$ for $exp(-\beta n\varepsilon_i) = p_i^\beta$.

from equations~\ref{equ:f83} and ~\ref{equ:f84},  we can get the entropy:
 
\begin{equation}
S(\overline{\varepsilon})=F(\beta)-\beta\frac{\mathrm d F(\beta)}{\mathrm d \beta }=-\beta^2 \frac{\mathrm d [F(\beta)/\beta]}{\mathrm d \beta }
\label{equ:f85}
\end{equation}

when $\beta = \beta_c, F(\beta_c)=0$, from equations~\ref{equ:f81}

\begin{equation}
\beta_c=D_0
\label{equ:f86}
\end{equation}

This is the fractal dimension that familiar to us. In addition, another quantity we are interested in is the escape index $ \delta $. From its definition, for large $n$,

\begin{equation}
\sum\limits l_i^d=\sum\limits_i exp(-n\varepsilon_i d)\sim exp(-n\delta)
\label{equ:f87}
\end{equation}

Here $d$ is the dimension of fractal embedded space. From equations~\ref{equ:f80}

\begin{equation}
\delta=-F(d)
\label{equ:f88}
\end{equation}

From equations~\ref{equ:f84}, ~\ref{equ:f85} and ~\ref{equ:f88},  it can be seen that as long as $F(\beta)$ is known, $\varepsilon, S(\varepsilon)$ and $\delta $ can be obtained. In summary, according to this thermodynamic description, the entropy function $S(\varepsilon)$ and the free energy $F(\beta)$ fully characterize the fractal topology, that is, the geometric characteristics.

\subsubsection{Thermodynamic form of fractal with measure}

The above thermodynamic forms can be generalized to the more general multi-scale fractal situation with probability measure distribution. Suppose that in the $n-th$ division, the $i-th$ sphere divided has a measure $P_i$, which satisfies the scaling relationship,

\begin{equation}
P_i \sim l_i^{\alpha_i}
\label{equ:f89}
\end{equation}

And also $ l_i=exp(-n\varepsilon)$

This is a distribution involving two scale indices $\varepsilon$ and $\alpha$. We count the number of balls whose scale index is between $\varepsilon \to \varepsilon+d\varepsilon $ and $\alpha \to \alpha+d\alpha$ is $ \Omega(\varepsilon,\alpha )d\varepsilon d\alpha$.  We expect for large n

\begin{equation}
\Omega(\varepsilon,\alpha )=exp[nQ(\varepsilon,\alpha )]
\label{equ:f90}
\end{equation}

$\Omega(\varepsilon,\alpha ) $ called the generalized entropy function. Now define the partition function:

\begin{equation}
\Gamma(q,\beta)=\sum\limits_ip_i^q l_i^\beta=\sum\limits_ip_i^q l_i^{-\tau}=\sum\limits_iexp[-n\varepsilon_i(\alpha_iq+\beta)]
\label{equ:f91}
\end{equation}

Corresponding free energy is defined as:

\begin{equation}
G(q,\beta)=\frac{1}{n} ln\Gamma(q,\beta)
\label{equ:f92}
\end{equation}

Obviously, the previous definition $Z(\beta)$ and $F(\beta)$ is a special case here

\begin{equation}
Z(\beta)=\Gamma(q=0,\beta)
\label{equ:f93}
\end{equation}

\begin{equation}
F(\beta)=G(q=0,\beta)
\label{equ:f94}
\end{equation}

Using equations~\ref{equ:f90} and~\ref{equ:f91}, $\Gamma(q,\beta)$ can be written to integral form:

\begin{equation}
\Gamma(q,\beta)=\int \mathrm{d}\varepsilon\int \mathrm{d}\alpha  exp\{n[Q(\varepsilon,\alpha )-(\alpha q+\beta )\varepsilon ]\}
\label{equ:f95}
\end{equation}

Combing equations~\ref{equ:f92}, according to the steepest descent method , we have:

\begin{equation}
G(q,\beta)=Q(\overline{\varepsilon},\overline{\alpha})-(\overline{\alpha} q+\beta )\overline{\varepsilon}
\label{equ:f96}
\end{equation}

where $\varepsilon$ and $\alpha$ gives the max value of $Q(\varepsilon,\alpha )-(\alpha q+\beta )\varepsilon  $, we have:

\begin{equation}
\left.\frac{\partial Q(\varepsilon,\alpha)}{\partial \varepsilon} |_{\varepsilon=\overline{\varepsilon}, \alpha=\overline{\alpha}}\right.=\overline{\alpha} q+\beta
\label{equ:f97}
\end{equation}

\begin{equation}
\left.\frac{\partial Q(\varepsilon,\alpha)}{\partial \alpha} |_{\varepsilon=\overline{\varepsilon}, \alpha=\overline{\alpha}}\right.=\overline{\varepsilon} q
\label{equ:f98}
\end{equation}

Or

\begin{equation}
\overline{\varepsilon}=-\frac{\partial}{\partial \beta}G(q,\beta)
\label{equ:f99}
\end{equation}

\begin{equation}
\overline{\varepsilon} \bullet \overline{\alpha}=-\frac{\partial}{\partial q}G(q,\beta)
\label{equ:f100}
\end{equation}

Thus:

\begin{equation}
Q(\overline{\varepsilon} , \overline{\alpha})=G(q,\beta)-q \frac{\partial G(q,\beta)}{\partial q}-\beta\frac{\partial G(q,\beta)}{\partial \beta}
\label{equ:f101}
\end{equation}

So as soon as the generalized free energy $G(q,\beta)$ is known, $\varepsilon$, $\alpha$ and entropy $Q(\varepsilon,\alpha)$ can be obtained.  Also let 

\begin{equation}
G(q,\beta=\beta_c(q))=0
\label{equ:f102}
\end{equation}

From

\begin{equation}
\Gamma(q,\beta=\beta_c(q))=\sum\limits_iP_i^q \bullet l_i^{\beta_c(q)}=1
\label{equ:f103}
\end{equation}

 Mass scale $\tau(q)$ in equation ~\ref{equ:f91} must satisfy: 

\begin{equation}
\tau(q)=(q-1)D_q
\label{equ:f104}
\end{equation}

From equations ~\ref{equ:f91} and  ~\ref{equ:f75}:

\begin{equation}
\beta_c(q)=-\tau(q)=(1-q)D_q=f(\alpha)-\alpha q
\label{equ:f105}
\end{equation}

$\beta_c(q)$ and $D-q$ can be obtained through the solution of equation ~\ref{equ:f102}

At the critical point $\beta = \beta_c$, from equations ~\ref{equ:f96},  ~\ref{equ:f97} and ~\ref{equ:f102}:

\begin{equation}
Q(\overline{\varepsilon}_c,  \overline{\alpha}_c) = \left.\frac{\partial Q(\varepsilon,\overline{\alpha}_c )}{\partial \varepsilon} |_{\varepsilon=\overline{\varepsilon}_c}\right. \bullet \overline{\varepsilon}_c
\label{equ:f106}
\end{equation}

let 

\begin{equation}
f(\overline{\alpha}_c) = \left.\frac{\partial Q(\varepsilon,\overline{\alpha}_c )}{\partial \varepsilon} |_{\varepsilon=\overline{\varepsilon}_c}\right.
\label{equ:f107}
\end{equation}

substitute ~\ref{equ:f107} into ~\ref{equ:f97} and ~\ref{equ:f98}, 

\begin{equation}
f(\overline{\alpha}_c) = \overline{\alpha}_c q +\beta_c(q)=\overline{\alpha}_c q -\tau(q)
\label{equ:f108}
\end{equation}

\begin{equation}
\left.\frac{\mathrm d f(\alpha)}{\mathrm d \alpha } |_{\alpha=\alpha_c}\right.=q
\label{equ:f109}
\end{equation}

equation ~\ref{equ:f108} give:

\begin{equation}
\overline{\alpha}_c = -\left.\frac{\mathrm d \beta_c(q)}{\mathrm d q}\right.
\label{equ:f110}
\end{equation}

Once $\beta_c(q)$ is obtained by solving equation ~\ref{equ:f102}, $\alpha_c$ and $f(\alpha_c)$ can be obtained from equation ~\ref{equ:f110} and equation ~\ref{equ:f108}. It is easy to prove that $f(\alpha_c)$ is the singularity spectral function of the multiscale fractal.

In summary, we see that the generalized free energy $G(q, \beta)$ plays a special role in the thermodynamics. This thermodynamic form of scale fractal gives a more comprehensive and rich description.

In some literatures ~\cite{Bohr1987,Feigenbaum1987,Bohr1987A,Tel}, other definitions equivalent to free energy are used and corresponding fractal thermodynamics are established. It can be proved that they are equivalent under certain circumstances.

\subsubsection{Fractal thermodynamic phase transition}

In the previous description, the following assumption is generally implied: in the entire value range of $q$, $-\infty <q < \infty$, when the partition function limit is $1 \to 0$, it can always be scaled to $\sum\limits_{i=1}^{N}P_i^q \simeq A(q)l^{\tau(q)}$ according to the definition of the literature . We call this type of fractal thermodynamic body's phase change a continuous phase change or a secondary phase change~\cite{Jensen,Artuso,Csorda1988,Csorda1989}.

However, for a certain system, this assumption may only be established in a certain region of $q$, while in other regions of q it may follow other types of non-power-law scaling. For example, the following two simplest forms:~\cite{Csorda1989}

\begin{equation}
\sum\limits_{i=1}^{N(n)}P_i^q \sim B(q)(ln\frac{1}{l})^{-\delta(q)}
\label{equ:f111}
\end{equation}

\begin{equation}
\sum\limits_{i=1}^{N(n)}P_i^q \sim B(q)C(q)^{-\frac{1}{l ^ { \delta (q)}}}
\label{equ:f112}
\end{equation}

Here $B(q)$,$C(q)$,$\delta(q)$  is a constant related to $q$.
In a box covering a fractal set, if there is at least one box, the probability on it as a function of box size $l$ will decay to zero more slowly than any power law, for example

\begin{equation}
P_s > \frac{1}{ln ^ \delta (1/l)}        ,    \delta>0
\label{equ:f113}
\end{equation}
At this time, for $q>l$, an upper bound of $D_q$ can be obtained.

\begin{equation}
0\leq D_q \leq \lim_{l\to 0}\frac{q}{1-q} \bullet \frac{ln P_s}{ln (1/l)} \leq - \lim_{l\to 0}\frac{q \delta}{1-q} \bullet \frac{lnln (1/l)}{ln (1/l)} =0
\label{equ:f114}
\end{equation}
In this way, if the probability decay of some boxes is slower than power law, then for $q>l$, $D_q$ is $0$, and the partition function can be scaled as ~\ref{equ:f111}. In this case, $f(\alpha)$ can take any value between $0$ and $D_1$ for $q>1$, $\alpha(q) =f(\alpha(q))=0 $ and $ q_c=1, \alpha(q=1) =f(\alpha(q=1))$,  so the phase change at $q_c = 1$ gives a straight line about $f(\alpha)$:

\begin{equation}
f(\alpha)=\alpha , \alpha \in [0, D_1]
\label{equ:f115}
\end{equation}

It means that only the value of $q$ less than $1$ in the relationship curve between the generalized dimension $D_q$ and $q$ can contribute to the reserved part of $f(\alpha)$. Similarly, if there is at least one box with a probability of decaying to zero faster than any power law, then for $q <0$, the partition function is dominated by the probability of these abnormally fast decay. For example, there is a box probability $P_f$

\begin{equation}
P_f<C^{-\frac{1}{l^\delta}}
\label{equ:f116}
\end{equation}
It can be inferred at this time: $q <0 $

\begin{equation}
D_q \geq \lim_{l\to 0}\frac{q}{1-q} \bullet \frac{ln P_f}{ln (1/l)} \geq - \lim_{l\to 0}\frac{-q}{1-q} \bullet \frac{lnC}{l^\delta ln (1/l)} =\infty
\label{equ:f117}
\end{equation}

For $f(\alpha)$, from mediocre inequality:

\begin{equation}
f(\alpha) \leq \alpha
\label{equ:f118}
\end{equation}

when $q<1$ have 

\begin{equation}
\alpha(q) \geq D_q
\label{equ:f119}
\end{equation}

Compare this formula with ~\ref{equ:f117}:

\begin{equation}
\infty=D_q \leq  \alpha(q) , q<0
\label{equ:f120}
\end{equation}

It shows that the negative $q$ value does not contribute to $f(\alpha)$, and the $f(\alpha)$ spectrum contains only the monotonically increasing part. In the case of ~\ref{equ:f116}, the partition function will be scaled according to ~\ref{equ:f112}. Usually ~\ref{equ:f111}, ~\ref{equ:f112} the two abnormal scale forms are limited to a subset of $q \in [-\infty, \infty]$, they define anomalous phases, in which $ D_q$ takes $0$ or $\infty$. The phase change occurs when approaching the boundary point $q_c$ of these terms. In contrast to the definition of the aforementioned secondary  phase transition, this type of phase transition is called a first-order phase transition.

\subsection{Multiscale fractal wavelet transform}

Although $ D_q$ and  $f(\alpha)$  are more refined than a single fractal dimension $ D_0$, they can characterize the local self-similarity of uneven fractal, but it can only provide statistical information of the fractal scale, but cannot completely describe the subtle local properties of the fractal and dynamic characteristics. It was found that the developed wavelet transform is a powerful tool to reveal the nature of fractal local scaling~\cite{Argoul1989A,Arneodo1988,Davis1994,Argonl1990}. Wavelet transform is a powerful mathematical tool in signal analysis~\cite{Combs,Danbechies,Argoul1989}. In frequency spectrum analysis, it can decompose any dimensional signal into independent contributions of time and frequency. At the same time, the information contained in the original signal is not lost. This decomposition method is carried out by selecting the appropriate wavelet and then translating and scaling. Therefore, it is equivalent to a mathematical microscope.It has a zoom and shift function. By examining the behavior of the system at different magnifications, one can further infer its dynamic roots. For these reasons, it is obviously very important for studying multi-scale fractals~\cite{Argoul1989}.

The key to constructing a mathematical microscope by wavelet transformation is to choose a suitable wavelet function

The wavelet $g_{a,b}(t)$ is a kind of function that can be constructed with a single function $g(t)$.

\begin{equation}
g_{a,b}(t)=A(a,b)g(t)=a^{-1/2}g(a^{-1}(t-b)) , t \in R
\label{equ:f121}
\end{equation}

Here $a$ and $b$ can be arbitrarily selected $(a,b \in R, a>0)$. The operator $A(a,b)$ represents the role of the affine group. The function that can be a candidate wavelet must satisfy the compatibility condition

\begin{equation}
C_g=2\pi \int \mathrm{d} \omega|g(\omega)|^2/\omega < \infty
\label{equ:f122}
\end{equation}

Here the wavelet $g \in L^2(R, \mathrm{d}x) \cap L^1(R, \mathrm{d}x)$. This condition is equivalent to $ \int \mathrm{d}t g(t) = 0$

There are three common wavelets that satisfy the above conditions.

\subsubsection{Morlet wavelet}

\begin{equation}
g_{\Omega}(t)= e^{i\Omega t}(e^{-t^2/2}-\sqrt{2}e^{-\Omega^2/4} e^{-t^2})
\label{equ:f123}
\end{equation}

\subsubsection{Piecewise constant wavelet}

\begin{equation}
g(t) = \left\{
  \begin{array}{ll}
    1 &   |t| < 1\\
    -1/2 &   1<|t| < 3\\
    0 &   |t|> 3
  \end{array}
\right.
\label{equ:f124}
\end{equation}

\subsubsection{Mexican hat}

The Mexican hat is the most commonly used wavelet form, which is actually a Gaussian function

\begin{equation}
g(t)= (1-t^2)e^{-t^2/2}
\label{equ:f125a}
\end{equation}

Its Fourier transform is:

\begin{equation}
g(\omega)= \omega^2e^{-\omega^2/2}
\label{equ:f125b}
\end{equation}

Because multi-scale fractal is to consider the local behavior of the system, therefore, using the mathematical tool of wavelet transform will help to test the local scale behavior and provide direct evidence for the multi-scale fractal structure of the system. 
First, we do not consider the divergence problem of multi-scale fractal spectrum, but it is universal to set the scaleability, that is, the local scale behavior does not cause scale. So if you define any point $x_0$

\begin{equation}
f_{x_0}(x)=f(x_0+x)-f(x_0)
\label{equ:f126}
\end{equation}

We have:

\begin{equation}
f_{x_0}(\lambda x)  \sim \lambda ^{\alpha(x_0)}  f_{x_0}(x)
\label{equ:f127}
\end{equation}

This behavior can be directly reflected by the following wavelet transform

\begin{equation}
T_g(\lambda a, x_0+\lambda b)= C_g^{-\frac{1}{2}} \int \mathrm{d} x ( \lambda /a)^{-\frac{1}{2}} g^*(( x - b)/a)f_{x_0}(\lambda x)
\label{equ:f128}
\end{equation}

Thus: 

\begin{equation}
T_g(\lambda a, x_0+\lambda b)=\lambda ^{\alpha(x_0)+1/2}T_g(a, x_0 + b)
\label{equ:f129}
\end{equation}

Equation ~\ref{equ:f129} shows that each singular value undergoes a wavelet transform to obtain a cone-shaped structure on the half-plane, so that the wavelet transform actually gives the spatial distribution of the multi-scale fractal singular values. In addition, in the "mathematical microscope" constructed by equations ~\ref{equ:f128}, ~\ref{equ:f129}, the parameter $a^{-1}$ is equivalent to the magnification, and $b$ is the position parameter. By transforming these two parameters, the local scale characteristics can be studied. Therefore, for any measure $\mu$, the wavelet transform can be defined as:

\begin{equation}
T_g(a, b)= a^{-n} \int g((x-b)/a) \mathrm{d} \mu (x)
\label{equ:f130}
\end{equation}

If measure $\mu$ satisfy

\begin{equation}
\mu (I( x_0, \lambda \varepsilon)) \sim \lambda ^{\alpha(x_0)} \mu(I(x_0,\varepsilon))
\label{equ:f131}
\end{equation}

We have:

\begin{equation}
T_g(\lambda a, x_0+\lambda b) \sim \lambda ^{\alpha(x_0)-n}T_g(a, x_0 + b)
\label{equ:f132}
\end{equation}

Generally, for $\lambda \in R$, the formula ~\ref{equ:f132} does not necessarily hold. The sufficient and necessary condition for the establishment of this formula is that $\lambda$ belongs to a complex sequence: $ \lambda_m \sim \beta^m , m \in Z$.  Therefore, the exponent $\alpha$ is also complex, so the scaling behavior of the corresponding fractal is oscillating~\cite{Holscheider,GoupillaudP,Farmer,Hanusse}. This periodic $ln\beta$ oscillation of the wavelet transformation is carried out around the slope $\alpha$, thus depicting a vivid picture of the spatial distribution of the index $\alpha$.

\section{Negative Fractal and Complex Fractal }

\subsection{Negative fractal dimension and its significance}
We are no stranger to the concept of negative dimensions. In the early 20s of last century, the distinguished Soviet mathematician Ulisson laid down the theory of dimensionality of topological space. According to this theory, we can reasonably agree: if and only if the topological space $X$ in the empty set, the dimension of $X$ is $-1$, ie $dimX = -1 \Leftrightarrow X =\Phi $. The question now is: can negative dimensions be generalized to fractal sets, can negative dimensions change continuously, that is, can negative scores be taken? If so, what is the meaning of negative fractal dimension? The answer to the first question is affirmative, that is, a negative dimension can also be defined for the fractal set, and the negative dimension can also be continuously evaluated, that is, it can be scored. The negative fractal dimension quantitatively describes the degree of empty collection. Before introducing the negative fractal dimension, we only studied the multifractal spectrum $f (\alpha)$ positive value $(f (\alpha)>0)$. But R. Blumenfeld ~\cite{Blumenfeld} and C. Meneveau et al. ~\cite{Meneveau} when studying the fractal characteristics of DLA and turbulence, found that the multifractal spectrum $f (\alpha)$ has negative values, after introducing the concept of negative fractal dimension, we can also study random multifractal when value $(f (\alpha)<0)$ at some $\alpha$ value. Mandelbrot ~\cite{Mandelbrot1990} believes that negative $f (\alpha)$ values ​​may contain richer physical connotations than positive $f (\alpha)$. He pointed out that negative $f (\alpha)$ plays the role of sampling variable in random fractals, which describes the expected fluctuations in samples of limited scale. There is a negative fractal dimension in turbulence and DLA, so the negative dimension can be further applied to the study of turbulence and DLA models.

\subsection{The formation mechanism of negative fractal dimension}

Let $S_1,S_2 $ be the two sets $S_1, S_2 \subset E^d $ in the $d$-dimensional Euclidean space $E^d$, and can be a normal set or a fractal set. The rest dimensions are $d-dimS_1, d-dimS_2$ respectively, then the intersection of $S_1$ and $S_2$  $S = S_1 \cap S_1 $ codimensional dimension is~\cite{Mandelbrot1990}:

\begin{equation}
\begin{aligned}
d-dim S &=d-dim S_1+ d- dim S_2 \\
dim S &= dim S_1+ dim S_2 - d
\end{aligned}
\label{equ:f133}
\end{equation}

Equation~\ref{equ:f133} is called the law of codimensionality. Generalizing ~\ref{equ:f133} to the fractal structure, we have the following rules of intersection:

\begin{equation}
D_c=D_A + D_B -d_T
\label{equ:f134}
\end{equation}

$D_A, D_B $ to represent the fractal dimensions of fractal $ A $ and $B$ respectively, and $d_T$ to represent the topological dimension of the space where fractal are located, $D_c$ is the fractal dimension of the intersection of the two fractals $ A $ and $B$

Although the co-dimensional law is simple, it is very useful. In some cases, it is difficult to observe the inside of a closed body, such as many works in the fields of geology, earthquake, geophysical prospecting, etc., so it is difficult to directly find its fractal dimension value. At this time, the law of co-dimension can be used to measure the fractal dimension of its internal physical quantity with some external features. At the same time, the negative fractal dimension can be obtained directly from the law of co-dimensionality, when $d-dimS_1+d-dimS_2 > d $, $ dim S $ is negative. 

\subsection{Complex fractal dimension}
The introduction of negative fractal dimension extends the concept of fractal dimension to the entire real number domain. Just as numbers have real numbers and complex numbers, do complex fractal dimensions also exist~\cite{Long,Pietronero1988A}?
From the introduction of fractal self-similarity, consider the correlation function of the occupancy number $n(r)$:

\begin{equation}
c(r)=<n(r_0)n(r_0+r)>_0
\label{equ:f135}
\end{equation}

This function defines how many occupied points are from the occupied point $r_0$ at $r$ (for simplicity, let $r$ be a quantity). Average value $<n(r_0)n(r_0+r)> $ is in terms of all possible choices of $r$. For random fractals, you can average all the patterns that can be achieved, and integrate ~\ref{equ:f135} to obtain the relationship between the average volume occupancy and the length. For a uniform fractal of $d$, we can have:

\begin{equation}
N(R)= \int_0^R \mathrm{d}r <n(r_0)n(r_0+r)>_0 \simeq R^d
\label{equ:f136}
\end{equation}

For the fractal structure with discrete scaling law, the power law has the property of periodic amplitude modulation. Based on this, the complex fractal dimension can be introduced, write ~\ref{equ:f136}  into the following generalized form:

\begin{equation}
N(R)= R_e\{R^{d+i\delta}\}=R^d \cos (\delta \bullet lnR)
\label{equ:f137}
\end{equation}

Obviously $\delta$ is closely related to the same period:

\begin{equation}
\delta= 2\pi/ln b_0 
\label{equ:f138}
\end{equation}

For the fractal of the continuous scaling law case,$b_0 \to \infty, \delta \to 0 $, Formulas  ~\ref{equ:f137} and  ~\ref{equ:f138} only the real fractal dimension $d$ is left, which is consistent with  ~\ref{equ:f136}.

\section{Fractal Space }

The application of fractal in various fields is largely restricted by the imperfect development of fractal theory itself. In the past, regular and smooth sets and functions were the main objects discussed in mathematical geometry, for this, the classical calculus method can be applied. Structures with irregular sets like fractal geometry are considered "morbid". In recent years, because fractals have been widely used in fields other than mathematics, mathematicians have begun to realize that “morbid” non-smooth sets like fractals can and are worth studying. The proposal of fractal space idea laid the foundation for the further development of fractal theory.

\subsection{Definition of Fractal Space}

Let$ (\, X,d)\,$ be a complete metric space. $H(\,X)\,$ denotes the space whose points are the compact subsets of $(\,X,d)\,$, $x \in X$ and $ B \in H(\,X)\,$ define

\begin{equation}
d(x,B)= min \{ d(x,y) : y \in B \}
\label{equ:f139}
\end{equation}

d(x,B) is called the distance from the point x to the set B. 

Let $A, B \in H(\,X)\,$ define

\begin{equation}
d(A,B)= max \{ d(x,B) : x \in A \}
\label{equ:f140}
\end{equation}

d(A,B) is called the distance from the set $A \in H(\,X)\,$  to the set $B \in H(\,X)\,$ . 

The Hausdorff distance between points A and B in $H(\,X)\,$ is defined by 

\begin{equation}
h(A,B)=d(A,B) \lor d(B,A)= max \{ d(A,B) ,  d(B,A)\}
\label{equ:f141}
\end{equation}

It can be proved that $ \{H(\,X)\,, h\}$ form a complete metric space and is called the space of Fractals~\cite{Barnsley}.

\subsection{Fractal Space Time}
\begin{figure}[H]
\begin{center}
\epsfig{file=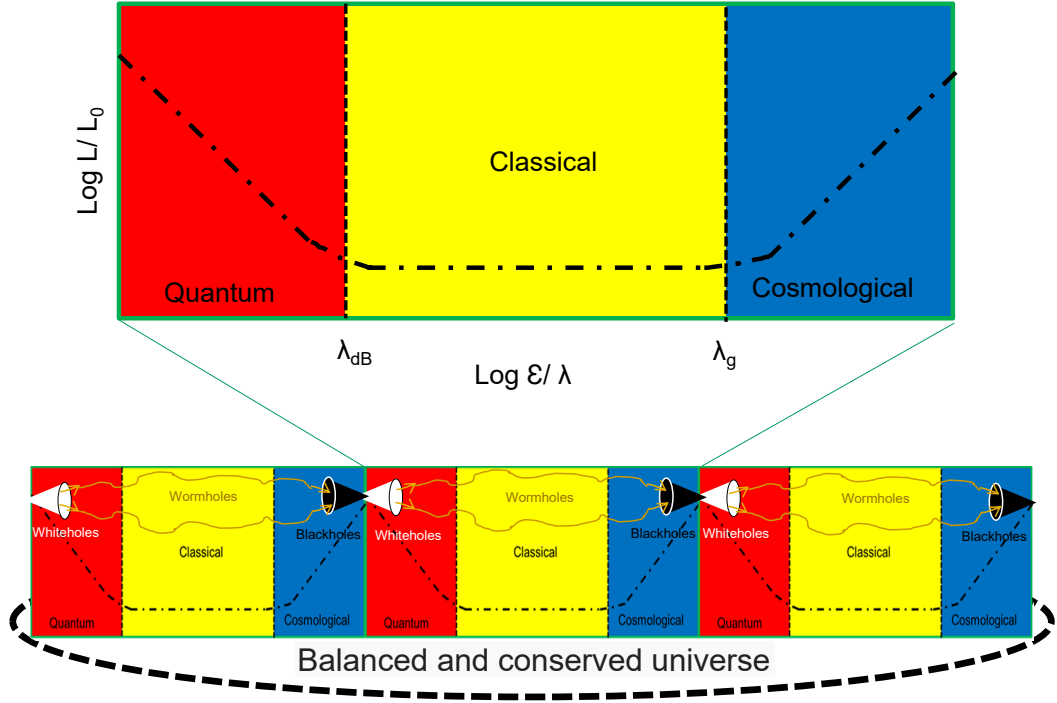,height=4.8in}
\caption{Balanced and conserved universe scheme}
\label{fig:figure3}
\end{center}
\end{figure}

If Einstein showed that space-time was curved, Nottale shows that it is not only curved, but also fractal. Nottale shows that a space which is continuous and non-differentiable is necessarily fractal~\cite{Nottale2011}. It means that such a space depends on scale. Mathematically, a fractal space-time is defined as a nondifferentiable generalization of Riemannian geometry~\cite{Nottale1994}. 
In the same way that general relativistic effects are not felt in a typical human life, the most radical effects of the fractality of spacetime appear only at the extreme limits of scales: micro scales or at cosmological scales. This approach therefore proposes to bridge not only the quantum and the classical, but also the classical and the cosmological, with fractal to non-fractal transitions (see Figure ~\ref{fig:figure3} top enlarged picture)~\cite{Nottale1996}. 

Based on this, we developed a balanced and conserved universe scheme and had been shown in Figure ~\ref{fig:figure3} bottom picture. It can be proved that the fractal to non-fractal transitions in dynamics is actually localization or super localization to dis-localization transition (see next Fractal Dynamics section for detail). The quantum range localization  to dis-localization transition had also been discussed in our previous works~\cite{Zhang2019A,Zhang2019B}. As shown in the Figure ~\ref{fig:figure3} bottom picture, the singularities that exist between the boundary of quantum and cosmological are whiteholes and blackholes ~\cite{Haggard,Dussaud}, wormholes  ~\cite{Garattini,Kirillov,Narahara}connect the whiteholes, blackholes everywhere and form the balanced and conserved universe. Future communication among universes will rely on the wormholes channels that connect to the whiteholes and blackholes everywhere in the universes since the information communication is instant and no time delay due to their singularity properties. We can use the small heavy molecular clusters finger print mass spectra change to detect the wormholes channels. The dimer, trimmer and tetrameter silver bromide clusters are the good candidates~\cite{Zhang2019B,Zhang1997A,Zhang2000A,Zhang2000B,Zhang2000C}, for the larger clusters, the resolution will get reduced~\cite{Zhang2019C,Zhang1991A,Zhang1992A,Zhang1995A,Zhang1994A}. 

In analogy with Einstein’s construction of general relativity of motion, which is based on the generalization of flat space-times to curved Riemannian geometry, it is suggested, in the framework of scale relativity, that a new generalization of the description of space-time is now needed, toward a still continuous but now nondifferentiable and fractal geometry (i.e., explicitly dependent on the scale of observation or measurement). New mathematical and physical tools are therefore developed in order to implement such a generalized description, which goes far beyond the standard view of differentiable manifolds. One writes the equations of motion in such a space-time as geodesics equations, under the
constraint of the principle of relativity of all scales in nature. To this purpose, covariant
derivatives are constructed that implement the various effects of the non differentiable and
fractal geometry. a Schrodinger-type equation (more generally a Pauli equation for spinors) is derived as an integral of the geodesic equation in a fractal space, then Klein-Gordon and Dirac equations in the case of a full fractal space-time. Nottale then briefly recall that gauge fields and gauge charges can also be constructed from a geometric reinterpretation of gauge transformations as scale transformations in fractal space-time~\cite{Nottale2011}.

\subsection{Manipulating Fractal Space Time}
Iterated function systems (IFS) are a simple, beautiful, deep, unified, natural mathematical framework via which affine geometry, topology, measure theory, dynamical systems, and probability can be meet and used. Here we discuss the fractal space time manipulating via the multi functions affine, projective and bilinear transformations and the probabilities associated with the transformations~\cite{Barnsley}.

\subsubsection{Affine transformation }
A transformation $w: \mathbb{R}^2 \to \mathbb{R}^2 $  of the form $w(𝑥_1,𝑥_2)=(𝑎+𝑏𝑥_1+𝑐𝑥_2,𝑑+𝑒𝑥_1+𝑓𝑥_2)$ where $𝑎,𝑏,𝑐,𝑑,𝑒,𝑓$ are real numbers, is called a two dimensional affine transformation. This definition can be easily generalized to large $N$ dimensional affine transformation. For the $xy$ Euclidean plane, affines have the form $(𝑎+𝑏𝑥+𝑐𝑦,𝑑+𝑒𝑥+𝑓𝑦)$. Affine functions require 6 numbers to be defined.  An invertible affine transformation is uniquely  determined by its action on three non collinear points, and conditions under which an affine transformation is a contraction in some metric.

\subsubsection{Projective and Mobius transformation}
A transformation $w: \mathbb{R}^2 \to \mathbb{R}^2 $  of the form $w(𝑥_1,𝑥_2)=(\frac{𝑎+𝑏𝑥_1+𝑐𝑥_2}{𝑔+ℎ𝑥_1+𝑖𝑥_2}, \frac{𝑑+𝑒𝑥_1+𝑓𝑥_2}{𝑔+ℎ𝑥_1+𝑖𝑥_2})$ where $𝑎,𝑏,𝑐,𝑑,𝑒,𝑓,𝑔,ℎ,𝑖$ are real numbers, is called a two dimensional projective transformation. For the $xy$ Euclidean plane, projective transformation have the form $(\frac{𝑎+𝑏𝑥+𝑐𝑦}{𝑔+ℎ𝑥+𝑖𝑦}, \frac{𝑑+𝑒𝑥+𝑓𝑦}{𝑔+ℎ𝑥+𝑖𝑦})$. Projective functions require 9 numbers to be defined.Similarly, we can define Mobius transformations on Riemann spheres in complex space. A transformation $f:\hat{\mathbb{C}}  \to \hat{\mathbb{C}}$  defined by: $f(z)=(\frac{𝑎z+𝑏}{cz+d}),$ where $𝑎,𝑏,𝑐,𝑑 \in \mathbb{C}, 𝑎𝑑-𝑏𝑐 \ne 0$ is called Mobius transformations on $\mathbb{C}$. If $ 𝑐 \ne 0$ then $ f(-𝑑/𝑐)= \infty$, and $ f(\infty)= 𝑎/𝑐$. if $ 𝑐 = 0$ then $ f(\infty)= \infty$.
Quantum fractals are patterns generated by iterated function systems, with place dependent probabilities, of Mobius transformations on spheres or on more general projective spaces~\cite{Jadczyk}. Platonic quantum fractals for a qubit was described in this work. Each vertex of Platonic solids serves as an attraction point. Thus a solid with $v$ vertices will define a SQIFS (Standard Quantum Iterated Function System) based on $v$ Mobius transformations~\cite{Jadczyk}.

\subsubsection{Bilinear transformation}
A transformation $w: \mathbb{R}^2 \to \mathbb{R}^2 $  of the form $w(𝑥_1,𝑥_2)=(𝑎+𝑏𝑥_1+𝑐𝑥_2+𝑑𝑥_1𝑥_2,𝑒+𝑓𝑥_1+𝑔𝑥_2+ℎ𝑥_1𝑥_2)$ where $𝑎,𝑏,𝑐,𝑑,𝑒,𝑓,𝑔,ℎ$ are real numbers, is called a two dimensional bilinear transformation. For the $xy$ Euclidean plane, bilinear transformation have the form $(𝑎+𝑏𝑥+𝑐𝑦+𝑑𝑥𝑦,𝑒+𝑓𝑥+𝑔𝑦+ℎ𝑥𝑦)$. Bilinear functions require 8 numbers to be defined.

\subsubsection{Probabilities}
An iterated function system with probabilities consists of an IFS $\{\mathbb{X}; w_1,w_2,\dotsc,w_N \}$ together with an ordered set of numbers $\{ p_1,p_2,\dotsc,p_N \}$, such that $\{ p_1+p_2+\dotsb+p_N =1\}$ and $ p_i>0$ for $\{i=1,2,\dotsc,N \}$. The probabilities allows you to change the probabilities $p_i$ assigned to the functions $w_i$ when playing the chaos game. The probabilities will be read as being ratios so the sum of all of the probabilities need not add to 1.

\subsubsection{Manipulating Fractal Space}
Fractal Space is manipulated by the manipulating space or codes that are composed of the real numbers of the transformation functions described above and the probabilities assigned to the functions. Here we use the IFS Generator program that developed by Brendan Harding and Michael Barnsley to demonstrate various  manipulating space or codes effects on the fractal space ~\cite{Barnsley}.

(1)Effects of transformation function types. 

In this study, we consider three  transformation function types discussed above and fix the the probability attached to these three transformation functions all the same 0.333. As shown in Figure 4, through the manipulating of the codes that composted by the real numbers of each function or the space formed by the codes, we can get exactly  the same  map dragon tile pattern in the fractal space. We conclude that the real pattern appeared in the fractal space independent to the type of transformations functions.

\begin{figure}[H]
\begin{center}
\epsfig{file=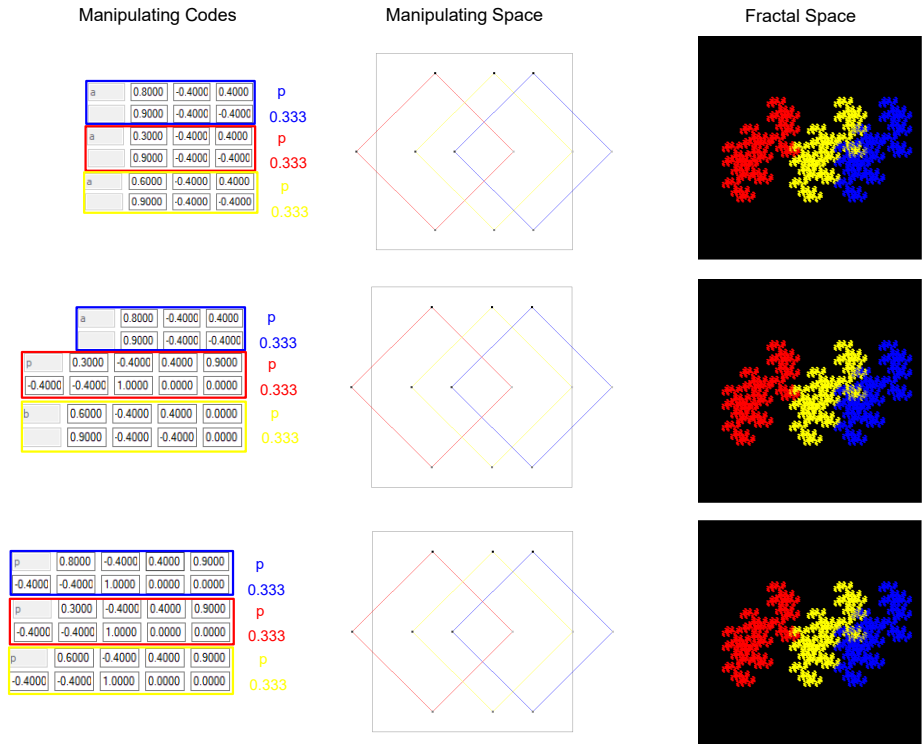,height=4.8in}
\caption{Effects of three types of transformation functions}
\label{fig:figure4}
\end{center}
\end{figure}

(2)Effects of probabilities of measurement.

Since the real pattern appeared in the fractal space independent to the type of transformations functions, we use affine transformation function for all three transformations and fix the manipulating code and space all the same in this study and vary the probability attached to each affine transformation function. The results had been shown in Figure 5. We found that the real pattern appeared in the fractal space dominated by the transformation function that has lowest probability attach to it.

\begin{figure}[H]
\begin{center}
\epsfig{file=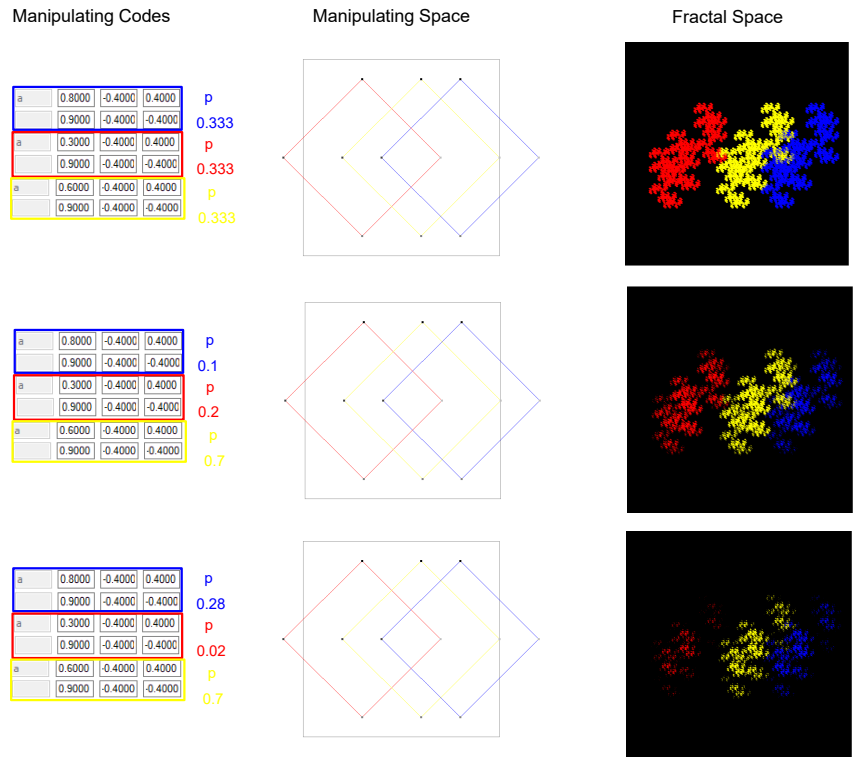,height=4.8in}
\caption{Effects of probabilities of measurement}
\label{fig:figure5}
\end{center}
\end{figure}

(3)Effects of manipulating codes.

Now let us use affine transformation function for all three transformations and fix the probability attached to each affine transformation function the same 0.333, just vary the manipulating codes and space, we found we can get all the various patterns that appeared in the nature in the fractal space as shown in Figure 6. This is amazing results!  This why we say iterated function systems (IFS) are a simple, beautiful, deep, unified, natural mathematical framework.

\begin{figure}[H]
\begin{center}
\epsfig{file=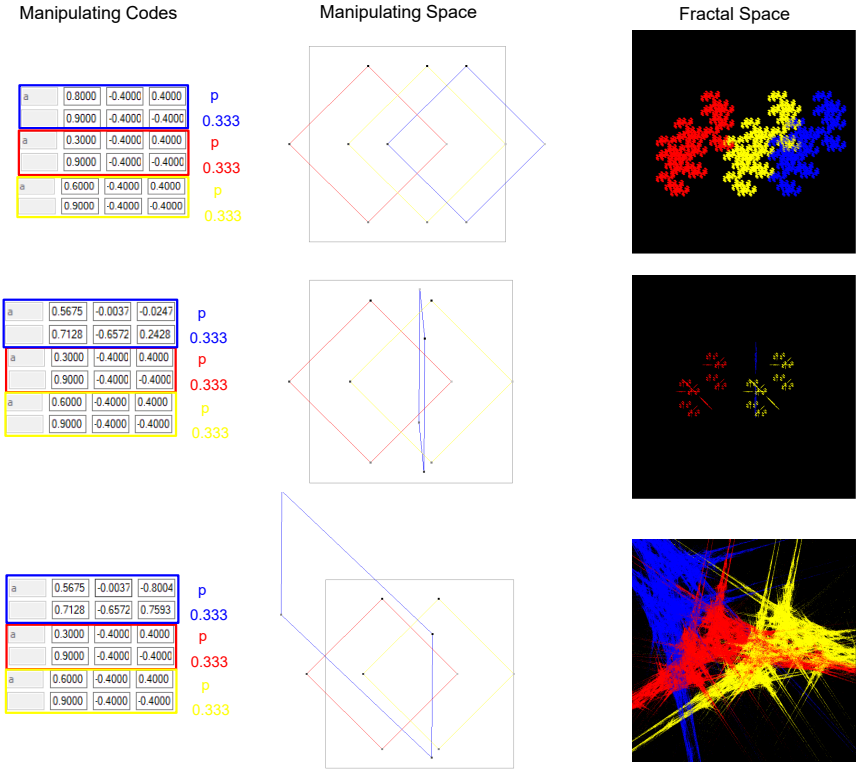,height=4.8in}
\caption{Effects of manipulating codes}
\label{fig:figure6}
\end{center}
\end{figure}

(4)Effects of hiding transformation functions.

A transformation function is called hiding transformation function if the probability attached to it is zero. To study the effects of hiding transformation functions, we consider two transformation systems: one ONLY have two affine transformation functions with the same probability 0.5 attached to each of them. The other have three affine transformation functions, two of the affine transformation functions with the same probability 0.5 attached to each of them, the third affine transformation function is an hiding transformation function with 0 probability attach to it, the patterns that appeared in the fractal space are compared in Figure 7. To our surprising, they are different! The one with hiding transformation function has discrete spot around the pattern in fractal space. The five large red and yellow spots are enlarged on purpose for easy view. Through the comparing studies of the real patterns appeared in the fractal space of these two designed systems, we know that whether there exists an invisible hand behind the system: the hiding transformation function.

\begin{figure}[H]
\begin{center}
\epsfig{file=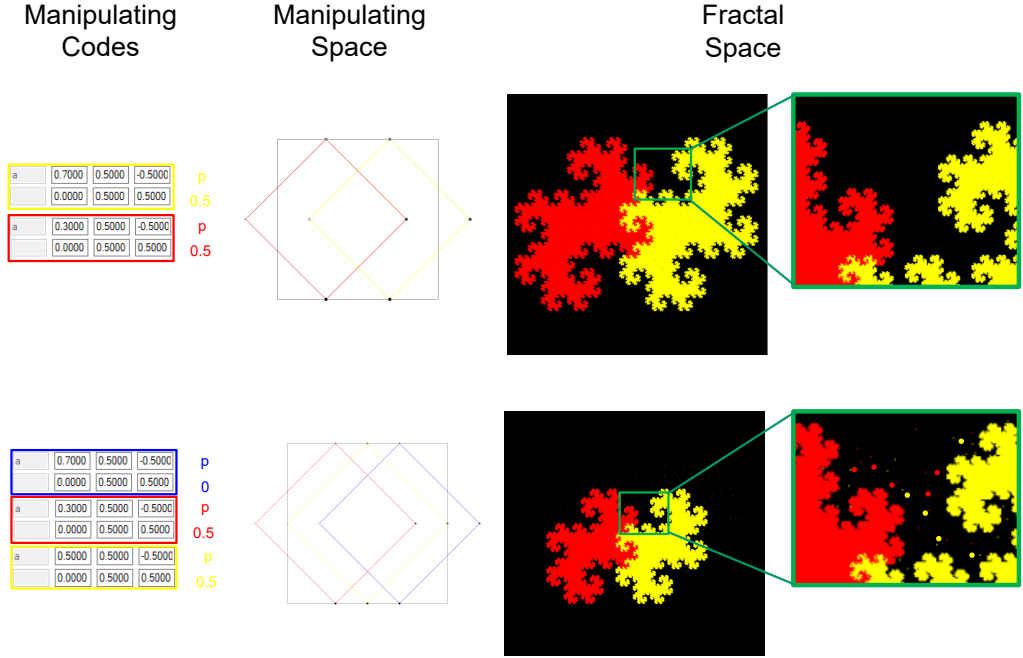,height=4.8in}
\caption{Effects of hiding transformation functions}
\label{fig:figure7}
\end{center}
\end{figure}

\section{Fractal Dynamics}

Earlier, we studied the static characteristics of fractal geometry and its characterization method. Now let's study the dynamic behavior of fractal structure: fractal dynamics. Broadly speaking, fractals can be defined as various element excitations on fractal structures. Vibration element excitations on the most studied fractal structures are currently being studied. That is to say, when solving the Hamiltonian equation, the potential energy term only considers the elastic potential, and then the scale-invariant boundary conditions of the fractal structure (Formula ~\ref{equ:f1}). Because the fractal nature is not like the periodic nature, waves vector $\textbf{k}$ can be introduced , the problem is transformed into reciprocal space, simplifying the solution. Therefore, there is need a new way to solve the vibration dynamic Hamilton on the fractal structure. At present, the most frequently used method is the vibration dynamic density method developed for disordered materials since the 1960s~\cite{Hori,Zheng,Zallen,Dai,Ziman,Elliott}. That is, first find the local density of states and then superimpose. Green function method and renormalization group method are commonly used to solve local state density. If there are some regrets to solve with this method, it is that the boundary conditions of the scale invariance (Formula ~\ref{equ:f1}) have not been fully reflected.
People have discovered and noticed the similarities and differences between the geometric localization of the vibration mode (fractal excitation) and the Anderson localization from the scattering (localized phonons)~\cite{Haydock,Peng,John},   this method is widely used by more and more researchers in the study of various element excitations on various fractal structures~\cite{Alexander,Aharony,Feng1985,Grest,Webman,Yakubo,Lam,Wohlman,Polatsek1988A,Courtens1987A,Tsujimi,
Peng1991,Li1990}. However, when Alexander Orbach put forward the concept of fracton~\cite{Alexander1982}, he used a completely different theoretical method — scale analysis, here we focuses on the introduction of this method.

\subsection{Scale analysis theory of dynamic behavior of fractal structure}

The actual fractal structure is always limited, and a transition between non-uniform and uniform material distribution will occur at a certain transition length $\xi$. Assuming that $\xi$ is the only characteristic length related to the physical problem, then according to the scale theory, the geometric dependence of matter at any length $L$ will be expressed by the ratio $L / \xi$. A typical intrinsic physical quantity $Q$ will change from the fractal behavior $Q \sim L^x $ when $L \ll \xi$ to the size independent behavior$ Q \sim \xi^x$ when $L \gg \xi$, and the dependence on $L$ and $\xi$ can usually be in the form of scale:

\begin{equation}
Q(L,\xi)=L^xF_Q(L/\xi)
\label{equ:f142}
\end{equation}
On a fractal network, the diffusion constant under random walking becomes related to the distance $r$. At this time, there is a characteristic dynamic index of elasticity or diffusion $\theta$, when $ r \ll \xi$, there are:

\begin{equation}
D(r)\propto r^{-\theta}
\label{equ:f143}
\end{equation}

The constant $\theta$ becomes zero when the structure approaches the Euclidean limit. Substituting the above formula into the diffusion equation  $<r^2(t)>=D(r)t$ gives:

\begin{equation}
<r^2(t)>\propto t^{2/(2+\theta)}
\label{equ:f144}
\end{equation}

In order to discuss the vibration properties of fractal structures, it is necessary to introduce another independent dimension, fracton dimension

\begin{equation}
\overline{d}=2D/(2+\theta)
\label{equ:f145}
\end{equation}

Thus:

\begin{equation}
<r^2(t)>\propto t^{\overline{d}/D}
\label{equ:f146}
\end{equation}

The vibration properties of the fractal structure can be characterized by the frequency $ \omega $. Because of the lack of wave vector $\textbf{k}$ as a good quantum number, only the characteristic length $\lambda$ of the mode can be used to establish the dispersion relationship $\omega = \omega(\lambda)$, and there is a transition between the uniform dispersion relationship and the fractal dispersion relationship between $\lambda \gg \xi$ and $\lambda \ll \xi$ . This transition occurs at $\omega_c = \omega(\xi) $. The concept of vibrational excitation from a long-wavelength (low-frequency) phonon to a short-wavelength (high-frequency) fractal on a fractal structure was first proposed by Alexander and Orbach~\cite{Alexander1982}. Aharony et al. also gave a scaling method~\cite{Aharony1985}.

The relationship between the transition length $\xi$ and the characteristic frequency $ \omega_c $ can be determined by the following formula, namely:

\begin{equation}
\omega_c \propto \xi^{-D / \overline{d}}=\xi^{-\frac{2+\theta}{2}}
\label{equ:f147}
\end{equation}

The fracton frequency is greater than $ \omega_c $ and the phonon frequency is less than $ \omega_c $. The correspondence between equations ~\ref{equ:f146} and ~\ref{equ:f147} is obvious, as long as the relationship between the diffusion problem and the vibration problem is noticed~\cite{Alexander1982}.

According to equation ~\ref{equ:f142}, we can write the scale form of the dispersion relationship:

\begin{equation}
\omega (\lambda,\xi) = \lambda ^{-\frac{2+\theta}{2}}f_\omega(\lambda/ \xi)
\label{equ:f148}
\end{equation}

Therefore, the dispersion relationship can be written in two different regions, namely at the phonon and fracton limits:

\begin{equation}
\omega_{ph} \sim c\lambda^{-1} \sim \xi^{-\theta/2}\lambda^{-1} , \lambda \gg \xi
\label{equ:f149}
\end{equation}

\begin{equation}
\omega_{fr} \sim \lambda^{-\frac{2+\theta}{2}}  , \lambda \ll \xi
\label{equ:f150}
\end{equation}

Where $c$ represents the sound velocity in the uniform region, $\lambda$ is the characteristic length of the mode, which is the wavelength under the phonon root limit, and the local length under the fracton limit. The two expressions are consistent to give the crossover frequency equation ~\ref{equ:f147}.

Now consider the density of states of vibration modes at different scales. We know that the density of phonon states is related to the Euclidean dimension, as:

\begin{equation}
N_{ph}(\omega) \sim \omega^{d-1}
\label{equ:f151}
\end{equation}

It can be proved that the density of fracton states is

\begin{equation}
N_{fr}(\omega) \sim \omega^{\overline{d}-1}
\label{equ:f152}
\end{equation}

\begin{figure}[H]
\begin{center}
\epsfig{file=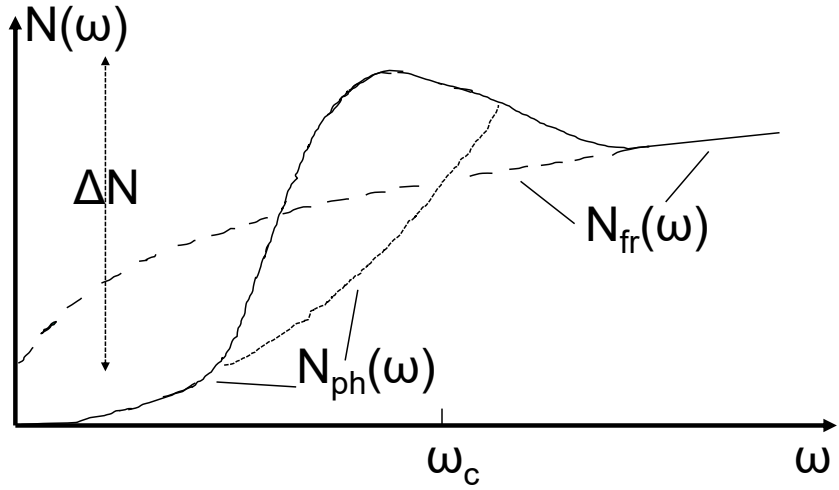,height=4.8in}
\caption{Diagram for the step $ \Delta N(\omega)$ in $N(\omega)$}
\label{fig:figure8}
\end{center}
\end{figure}

Among them, $\overline{d}$is the fracton dimension defined by formula  ~\ref{equ:f145}, which is introduced for accurate calculation of the number of frequency space modes, and is hereinafter referred to as spectral dimension.
At a finite $\xi$, the transition from $N_{ph}(\omega) (\omega \ll \omega_c) $ to $N_{fr}(\omega) (\omega \gg \omega_c) $ can be expected. Comparing equations ~\ref{equ:f151} and ~\ref{equ:f152}, since $ \overline{d} <d$, the low frequency region $N_{ph}$ is much smaller than$N_{fr}$. In order to ensure that the frequency summation of $N (\omega)$ is fixed, $ N(\omega)> N_{fr}(\omega)$ must be made at the crossing frequency $\omega_c$, that is, near the $\omega_c$, there is a step $ \Delta N(\omega)$ in $N(\omega)$, the step size is close to $ \omega_c^{\overline{d}-1}$, as shown in Figure ~\ref{fig:figure8}

As the development of the fracton concept, Alexander et al.~\cite{Alexander1985} pointed out that this high-frequency localized quantum vibration will cause relaxation due to the emission and absorption of localized electrons. The relaxation rate is calculated by the probability density. On the fractal network, the traditional electron-lattice interaction will be severely corrected. Two cases can be distinguished, local electron-single fracton interaction and local electron-double fracton interaction. To this end, we must explicitly consider the spatial extent of the wave function, and it is necessary to introduce a super-local index (also called the fourth dimension) $ d_\Phi$, used to determine the range of the local fractal wavelet function.

As another development of the fracton concept, Alexander et al.~\cite{Alexander1986} also considered that the interaction between phonons and fractons can exist in the following two processes and inverse processes: two phonons combined into a fracton, a phonon combined with a fracton Into another fracton. The former is important for the platform characteristics of the relationship between the thermal conductivity $\kappa$ of disordered solids and temperature, while the latter is important for $\kappa$ in the upper platform region linear temperature relationship.

Now we look at the distribution of relaxation time~\cite{Aharony}. In a sufficiently uniform region, the phonon state can be described by the wavelength $\lambda$ and frequency $\omega = c/\lambda $ plane wave, when there is a weak disorder, the pure plane wave will decay with time by elastic scattering into other waves. The golden rule can be used to derive the survival time.

\begin{equation}
1/\tau(\omega)=N(\omega)|V|^2
\label{equ:f153}
\end{equation}

$V$ is matrix elements that transform from the initial plane wave to other states at the same frequency. Using the perturbation Hamiltonian $ \frac{1}{2}\sum\limits_{ij} k_{ij}(\varphi_i-\varphi_j)^2 $, $\varphi_i$ is the displacement of the ith atom, $k_{ij}$ is the coupling constant, we can derive the corresponding survival time of the weak scattering caused by disorder:

\begin{equation}
1/\tau(\omega)\sim \omega^2 N_{ph}(\omega)\sim \omega ^{d+1} \omega_c^{-d}=\omega \bigg(\frac{\omega}{\omega_c}\bigg)^d
\label{equ:f154}
\end{equation}

It's Rayleigh's law. It can be written using the standard scale form ~\ref{equ:f142}

\begin{equation}
1/\tau(\omega)= \omega f_{\tau}(\omega/\omega_c)
\label{equ:f155}
\end{equation}

Extrapolating to the fracton region and requiring $f_{\tau}(\omega/\omega_c)$ to become a constant, there is:

\begin{equation}
1/\tau(\omega) \sim  \omega 
\label{equ:f156}
\end{equation}

\subsection{Computer simulation of dynamic behavior of fractal structure}

The scale analysis theory gives many interesting results, however, they need to be further computerized and experimentally verified. Percolation cluster is a good and commonly used example for computer simulation verification. Yakubo and Nakayama~\cite{Yakubo} and  Li etal ~\cite{Li1990} conducted a numerical study on the state density of the percolation clusters. They all confirmed the power exponential behavior between the fracton frequencies of the state density and crossed into the Debye phonon type spectrum in the low frequency region; and confirmed that the fractons are local states. But there is a disagreement on whether the fractal is hyperlocal. 
The results of Yakubo and Nakayama support hyperlocalization, while the results of Li and Vries~\cite{Vries} are negative.
The calculation results of the vibration spectrum on other fractal structures also confirm that the fractal is a local state and the phonon to fracton transition~\cite{Webman,Peng1991}. But there are still different opinions on the form of the transit~\cite{Webman,Peng1991,Derrida,Orbach,Pynn1985,Pynn1987}.

There are no reports of computer simulation experiments on the results of scale analysis theory about fracton life.

\subsection{Experimental study on dynamics of fractal structure}

The dynamic behavior of many fractal structures derived from scaling theory has been tested directly or indirectly on various fractal objects through various experimental methods~\cite{Alexander1983,Kelham,Zaitlin,Oliveira}.

The main characteristics of low-temperature specific heat and thermal conductivity of early amorphous materials can be said to indirectly support the description of fractal.
Ultrasonic attenuation experiments ~\cite{Page} neutron scattering experiments~\cite{Buchenau,Rosenberg,Freltoft1987,Freltoft1986,Fontana} for possible fractal structures in disordered materials; a series of inelastic scattering spectrum experiments on the fractal structures~\cite{Yabubo,Countens1987A,Vacher,Pelous,Calemciuk,Courtens1987,Courtens1989,Polatsek1988,Tsujimk,Boukenter} also confirmed the existence of fracton and phonon to fracton transition.

Uemura and Bigeneau  ~\cite{Uemura1986,Uemura1987} also used high-resolution inelastic neutron scattering experiments to confirm the existence of the transition from the magnetic oscillator to the fracton in the three-dimensional random dilution antiferromagnet $(Mn_xZn_{1-x}F_2)$

All in all, with the deepening of people's understanding of fractal structure and its properties, the dynamic behavior of fractal structure will be more and more valued by people. There is no doubt about this ~\cite{Feng1991}.

\section{Discussion of Applications of Fractals in Integrated Circuits}

\begin{figure}[H]
\begin{center}
\epsfig{file=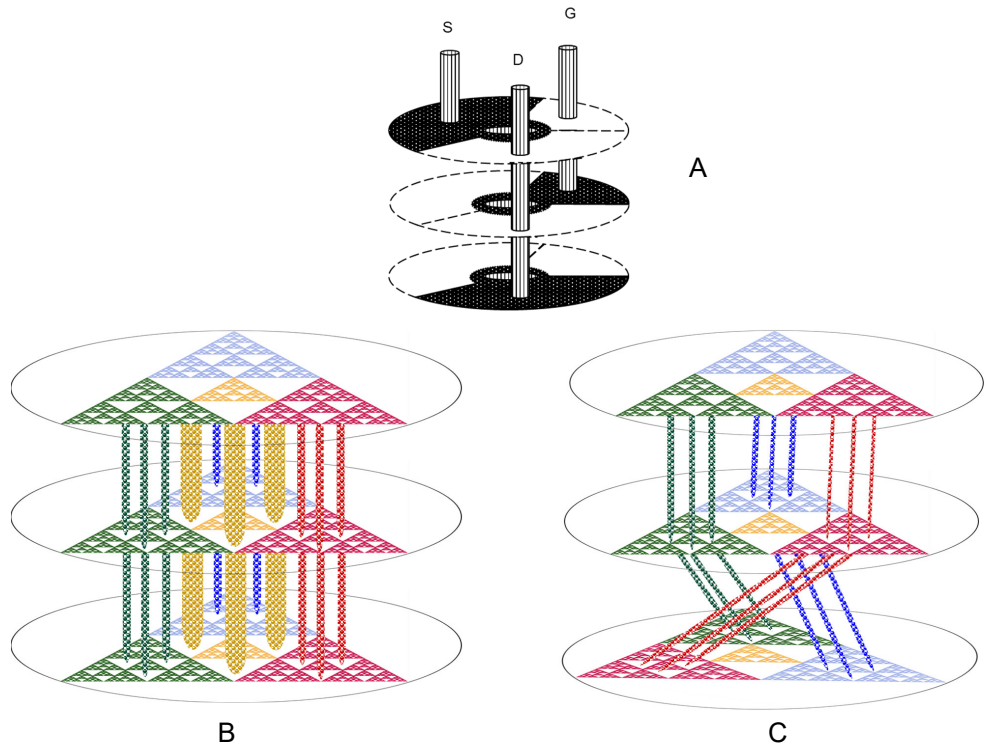,height=4.8in}
\caption{Device diagram for IC interconnections}
\label{fig:figure9}
\end{center}
\end{figure}

Integrated circuits have various device architectures such as Vertical Field Effect Transistor (VFET)~\cite{Zhang2018B}, Vertical Turnal Field Effect Transistor(VTFET)~\cite{Zhang2018C,Zhang2016A}, Nanosheet Transistor~\cite{Zhang2017B, Zhang2018D,Zhang2017C}, Photonic Integrated Circuits~\cite{Zhang2015B, Zhang2019D}, Bio sensor~\cite{Zhang2017D}, 3D stack~\cite{Zhang2013A}, 3D package~\cite{Zhang2019E}, 3D cooling~\cite{Zhang2014A}, Radio Frequence Integrated Circuits~\cite{Zhang2019F}, Analog Integrated Circuits~\cite{Zhang2019F}, Logic Integrated Circuits~\cite{Zhang2019G}, Input/Output devices (I/O)~\cite{Zhang2017E}, Fully Deplete Silicon on Insulator device (FDSOI)~\cite{Zhang2018E}, Dynamic Random Access Memory (DRAM)~\cite{Zhang2015C}, Static Random Access Memory(SRAM)~\cite{Zhang2017F},  Phase Change Memory(PCM)~\cite{Zhang2015D}, Magnetic Random Access Memory (MRAM)~\cite{Zhang2018F}, Resistive Random Access Memory (RRAM)~\cite{Zhang2019H}, Vacuum transistor~\cite{Zhang2017G}, Thin Film Transistor~\cite{Zhang2018G},  Crack Stop~\cite{Zhang2017H}, Efuse~\cite{Zhang2016B}, Electical Static Diode(ESD)~\cite{Zhang2015E}, SiC device~\cite{Zhang2015F}, High Electron Mobility Transistor (HEMT)~\cite{Zhang2015G} and micro-electromechanical system (MEMS) devices~\cite{Zhang2016C} etc. 

As technology node continues to shrink, high density of interconnections put into a limited area becomes extreme challenge for integrated circuits. To solve this issue, in our previous work, we proposed a novel vertical interconnect structure as shown in Figure~\ref{fig:figure9}A~\cite{Zhang2016D}. This still not solve the problem if we want to put extreme large amount of connections within a limited space. In nature, fractals have the properties to have infinite areas within a limited space theoretically. Figure ~\ref{fig:figure9}B we proposed a novel interconnection structures within triangle mania fractals. In the picture, we only show the first and second order connections, but theoretically this kind of orders can be infinities,thus the connections can also be infinities. The other advantage for these kinds of interconnections are that they can be grouped into different functional groups easily, for example, the red can for heat connection,the blue can for optical connection, the green can for magnetic connection and the yellow can for electrical connection, this is very important for the future new concept of transistors we proposed previously~\cite{Zhang2019B}. Figure ~\ref{fig:figure9}C shows that this kind of connection can also be twisted in an arbitrary angles to meet different functional requirements.

\section{Conclusion}
There are three important types of structural properties that remain unchanged under the structural transformation of condensed matter physics and chemistry. They are the properties that remain unchanged under the structural periodic transformation-periodic properties. The properties that remain unchanged under the structural multi scale transformation-fractal properties. The properties that remain unchanged under the structural continuous deformation transformation-topological properties. For periodic properties,People have done quite in-depth research,the whole set of concepts, theories and research methods has been widely used.For fractal properties, it was first proposed by B. Mandbrot in 1975, and he named the structures with fractal properties Fractal. Fractal dimension was introduced to quantitatively describe the geometric features of the fractal structure.  Establishing the geometric theory of fractal structures.This theory has been widely used in many disciplines such as physics, chemistry, biology, and geosciences. Which opened up a whole new field of research. Although topological properties have been studied in the mathematical world for more than a century, the application of topological properties in physics and chemistry is only a matter of recent decades.If the scientific historian Kuhn's paradigm theory on the process of scientific development is used to illustrate, it can be considered that the study of periodic nature has established a formal paradigm, while the study of fractal and topological properties is just before the establishment of the paradigm, and the discipline is not yet mature. At this time, different viewpoints can coexist, and knowledge is continuously accumulated, but there is no coherent theoretical system.In this paper, we have described some important methods used so far to characterize the fractal properties, including the theoretical method of calculating the fractal dimension, the renormalization group method, and the experimental method of measuring the fractal dimension. Multiscale fractal theory method, thermodynamic representation form and phase change of multiscale fractal, and wavelet transform of multiscale fractal. The  development of the fractal concepts is briefly introduced: negative fractal dimension, complex fractal dimension and fractal space time. New concepts such as balanced and conserved universe,the wormholes connection to the whiteholes and blackholes for universes communication, quantum fractals,  platonic quantum fractals for a qubit, new manipulating fractal space time effects such as transformation function types, probabilities of measurement,manipulating codes,and hiding transformation functions are also discussed.  In addition, we discussed the use of scale analysis theory to stimulate the elements on the fractal structure: the research on the dynamics of fractal structure and the corresponding computer simulation and experimental research. The novel applications of fractals in integrated circuits are also discussed in this paper.

\end{document}